\documentclass[aip,reprint,a4paper,twocolumn,english]{revtex4-1}
\usepackage[T1]{fontenc}
\usepackage[utf8]{inputenc}
\usepackage[english]{babel}
\usepackage{mathtools}
\usepackage{amsmath}
\usepackage{amssymb}
\usepackage{graphicx}
\usepackage{comment}
\usepackage{esint}
\usepackage{nicefrac}
\usepackage{dsfont}
\usepackage{bm}
\usepackage{hyperref}
\usepackage[linesnumbered,ruled,lined]{algorithm2e}
\usepackage{xcolor}
\usepackage[percent]{overpic}
\usepackage{siunitx}

\usepackage[standard,thmmarks,amsmath,hyperref]{ntheorem}

\renewtheorem{theorem}[z]{Theorem}
\renewtheorem{proposition}[z]{Proposition}
\renewtheorem{lemma}[z]{Lemma}
\renewtheorem{corollary}[z]{Corollary}
\renewtheorem{definition}[z]{Definition}
\renewtheorem{example}[z]{Example}
\renewtheorem{remark}[z]{Remark}

\newcommand{\bfR}{\mathbf{R}}
\newcommand{\bfr}{\mathbf{r}}
\newcommand{\id}{\mathrm{d}}
\newcommand{\eps}{\varepsilon}
\DeclareMathOperator{\var}{var}

\DeclareMathOperator{\dist}{dist}
\newcommand{\CC}{\mathcal{C}}
\newcommand{\CF}{\mathcal{F}}
\newcommand{\TC}{\text{C}}
\newcommand{\TF}{\text{F}}
\newcommand{\TI}{\text{I}}

\newcommand{\TA}{\text{A}}
\newcommand{\Var}{\operatorname{Var}}

\newcommand{\XX}{\mathcal{X}}
\newcommand{\sigS}{\sigma_{\rm S}}
\newcommand{\sigB}{\sigma_{\rm B}}
\newcommand{\muS}{\mu_{\rm S}}
\newcommand{\muB}{\mu_{\rm B}}
\newcommand{\etaS}{\eta_{\rm S}^{\rm r}}

\newcommand{\beq}{\begin{equation}}
\newcommand{\eeq}{\end{equation}}

\newcommand\numberthis{\addtocounter{equation}{1}\tag{\theequation}}

\begin{document}

\title{Multilevel simulation of hard-sphere mixtures}
\author{Paul~B.~Rohrbach}
\email{pbr28@cam.ac.uk}
\affiliation{Department of Applied Mathematics and Theoretical Physics, University of Cambridge, Wilberforce Road, Cambridge CB3 0WA, United Kingdom}
\author{Hideki~Kobayashi}
\affiliation{Institute for Theoretical Physics, Georg-August University Göttingen, Germany}
\author{Robert~Scheichl}
\affiliation{Institute for Applied Mathematics, Heidelberg University, Im Neuenheimer Feld 205, 69120 Heidelberg, Germany}
\affiliation{Department of Mathematical Sciences, University of Bath, Claverton Down, Bath BA2 7AY, United Kingdom}
\author{Nigel~B.~Wilding}
\affiliation{H.H.~Wills Physics Laboratory, University of Bristol, Royal Fort, Bristol BS8 1TL, United Kingdom}
\author{Robert~L.~Jack}
\affiliation{Department of Applied Mathematics and Theoretical Physics, University of Cambridge, Wilberforce Road, Cambridge CB3 0WA, United Kingdom}
\affiliation{Yusuf Hamied Department of Chemistry, University of Cambridge, Lensfield Road, Cambridge CB2 1EW, United Kingdom}

\date{\today}

\begin{abstract}
    We present a multilevel Monte Carlo simulation method for analysing multi-scale physical systems via a hierarchy of coarse-grained representations, to obtain numerically-exact results, at the most detailed level.
    We apply the method to a mixture of size-asymmetric hard spheres, in the grand canonical ensemble.  A three-level version of the method is compared with a previously-studied two-level version.  The extra level interpolates between the full mixture and a coarse-grained description where only the large particles are present -- this is achieved by restricting the small particles to regions close to the large ones.
     The three-level method improves the performance of the estimator, at fixed computational cost.
    We analyse the asymptotic variance of the estimator, and discuss the mechanisms for the improved performance.
\end{abstract}

\maketitle

\section{Introduction}

A key challenge in the molecular simulation of soft matter is posed by the separation of length-scales between its microscopic description and the existence or emergence of mesoscopic structure.
In such cases, one often relies on coarse-grained (CG) descriptions of the system that (approximately) integrate out microscopic degrees of freedom \cite{likos2001effective,noid2008multiscale,peter2009multiscale}, to yield a tractable simplified model.  Examples in the soft-matter context include polymers~\cite{Praprotnik2007,Pierleoni2007},  biomolecules\cite{Ouldridge2011,Mladek2013,Pak2018}, and colloidal systems~\cite{asakura1954interaction,roth2000depletion}.   Such CG descriptions are essential for multi-scale modelling approaches~\cite{Karplus2014,Warshel2014}.
However, they are not usually exact, and the associated coarse-graining errors are often difficult to assess.

Such CG models have been studied extensively in colloidal systems with depletion interactions~\cite{asakura1954interaction,Lekkerkerker1992,Poon2002}.
The typical example is a mixture of relatively large colloidal particles with much smaller non-adsorbing polymers which generate effective attractions between the colloids.   
This can drive de-mixing, crystallisation, or gelation, depending on the context.
Model systems in this context include the Asakura-Oosawa (AO) model \cite{asakura1954interaction} where the CG model can even be exact, if the disparity between colloid and polymer radii is large enough.
The theoretical tractability of the AO model arises from a simplified modelling assumption, that polymer particles act as spheres that can interpenetrate.

Alternatively, one may consider a mixture where both the colloids and the depletant are modelled as hard spheres.  
From a theoretical perspective, this is an interesting model in its own right as it undergoes a fluid-fluid de-mixing phase separation for large enough size-disparities and concentrations~\cite{biben1991phase,dijkstra1999phase,roth2000depletion,kobayashi2021critical}.
This happens despite the lack of attractive forces between the particles in the model, and can be attributed to geometric packing effects of the big and small spheres.

Direct simulation of such mixtures is very challenging, because of the large number of small particles.
Accurate CG models are available in this context too~\cite{roth2000depletion}, but the CG representations are not exact: 
their errors can be detected by accurate computer simulation of the full (FG) mixtures.
Hence, such models are natural testing grounds for theories and simulation methods associated with coarse-graining.

In this context, we recently developed a method \cite{kobayashi2019correction,kobayashi2021critical} that links a CG description with the underlying fine-grained (FG) description.
We call this the \emph{two-level} method, because the CG and FG models describe the same system, with different levels of detail.
The method was validated by computations on the AO system~\cite{kobayashi2019correction}, where it provided numerically-exact results for the FG model, even in the regime where the CG description is not quantitatively accurate.
The methodology was also applied to the hard sphere mixture~\cite{kobayashi2021critical}, where it provided a quantitative analysis of the critical point associated with de-mixing of the large and small particles.

These previous results rely on the idea that properties of the FG model can be estimated in terms of some CG quantity, with an additive correction that accounts for the coarse-graining error.
This is an importance sampling method, familiar in equilibrium statistical mechanics from free-energy perturbation theory\cite{zwanzig1954high}, which involves reweighting between two thermodynamic ensembles.
In the present context, the reweighting factors depend on the free energy of the small spheres, computed in a system where the large particles are held fixed.
This free energy can be estimated by an annealing process based on Jarzynski's equality~\cite{jarzynski1997nonequilibrium,crooks2000path,neal2001annealed} that slowly introduces small particles to fixed CG configurations. 

In this paper, we present an extension of the two-level method that incorporates additional intermediate levels to improve the overall performance.
Specifically, we introduce a step in the annealing process where small particles are partially inserted in regions close to big particles.
Before finishing the small-particle insertion, we then replace weighted sets of configurations with unweighted ones, duplicating configurations with large weight and deleting ones with low weight.
This resampling step allows us to make optimal use of the information available at the intermediate stage, focusing our subsequent computations on configurations that matter.

This general approach fits in the framework of sequential Monte Carlo (SMC) \cite{gordon1993novel,del2004feynman,doucet2009tutorial}.
Such algorithmic ideas have been successfully applied in applications across disciplines under various names, including population Monte Carlo\cite{iba2001population} or the go-with-the-winners strategy\cite{grassberger2002go}.
Examples in computational physics include the pruned-enriched Rosenbluth method for polymers\cite{hsu2011review}, the cloning method for rare events\cite{giardina2011simulating}, and diffusion quantum Monte Carlo\cite{reynolds1982fixed}.
We combine the SMC method with an additional variance reduction strategy.
Instead of estimating the FG average directly, we combine a CG estimate with estimates of subsequent level differences, using the previous levels as control variate.
This is the idea behind multilevel Monte Carlo methods\cite{giles2008multilevel,hoang2013complexity,dodwell2015hierarchical}.
The combination of a difference estimate with SMC has been previously investigated for example in Refs.~\onlinecite{jasra2017multilevel,beskos2017multilevel,del2017multilevel}.
As in Ref.~\onlinecite{kobayashi2021critical}, we develop a general method alongside its application to highly size-asymmetric binary hard-sphere mixtures, which provide a challenging but well understood example to benchmark our algorithm.

This paper is organised as follows:
In Section \ref{sec:HSMixture}, we introduce the hard-sphere mixture model.
Then in Section \ref{sec:method}, we summarise the setup of the two-level method before presenting our extension to three (or more) levels.
The three-level method requires an intermediate level for the hard-sphere mixture, whose details we discuss in Section \ref{sec:IntermediateLevel}.
In Section \ref{sec:NumericalResults}, we present a numerical test of the method and compare its performance against the two-level method, and in Section \ref{sec:ConvergenceResults} we present convergence results.
We conclude in Section \ref{sec:conclusions}.

\section{Hard-sphere mixture} \label{sec:HSMixture}

Throughout this work, we illustrate the multilevel method with an example system, which is a mixture of large and small hard spheres at size ratio $10 : 1$. 
This system is challenging for simulation because the big particles may display interesting collective behaviour (in particular, a critical point), but the dominant computational cost for simulation comes from the large number of small particles.  

However, despite our focus on this single example, we emphasise that the multilevel method is presented in a general way, which should be applicable also in other systems with a separation of length scales.

\subsection{Hard sphere mixture}

\begin{figure*}
    \includegraphics[width=\textwidth]{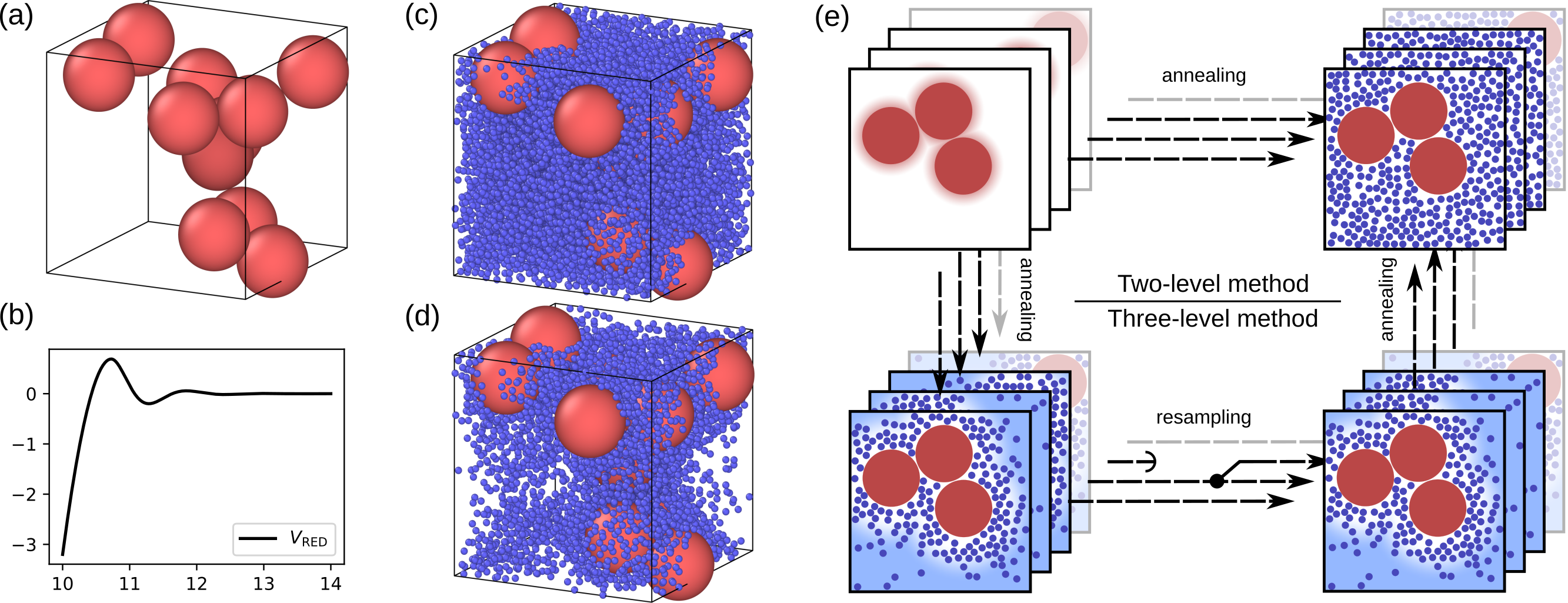}
    \caption{
        An overview of the levels of the example system from Section \ref{sec:example} and the structure of the three-level method.
        (a) A sample of the CG model $p_{\TC}$ with $N = 11$ big particles.
        (b) The two-body potential $V_{\text{RED}}$ used for the CG model.
        (c) A sample of the full FG model $p_{\TF}$ which has the big particle configuration as (a). This system contains $n = 8842$ small particles.
        (d) A sample of the partially inserted model $p_{\TI}$ used in the three-level algorithm. The small particles are primarily inserted around big particles, reducing the number to $n = 4473$.
        (e) A sketch of the two- and three-level method: starting with a population of CG configurations, we can directly compute importance weights by simulating an annealing process introducing the small particles (upper path, two-level method).
        Alternatively, we introduce a partially-inserted intermediate level where we interrupt this annealing process and resample to boost relevant configurations (lower path, three-level method).
    }
    \label{fig:OverviewMethod}
\end{figure*}

The example system is a mixture of big and small particles, whose diameters are $\sigB$ and $\sigS$ respectively. 
We consider a periodic box $[0,L]^3$ of linear size $L$ and we work in the grand canonical ensemble.
(This choice is particularly relevant for analysis of de-mixing, where the number density of large particles is a suitable order parameter\cite{bruce2003computational}.)

In a given configuration, the numbers of big and small particles are $N$ and $n$ respectively; the position of the $i$th big particles is $\bfR_i$ while the position of the $j$th small particle is $\bfr_j$.
We denote the configurations of big and small particles by $\CC = (N; \bfR_1, \dots, \bfR_N)$ and $\CF = (n; \bfr_1, \dots, \bfr_n)$ respectively, and the full configuration is denoted $\XX=(\CC, \CF)$.%

Since the particles are hard, the temperature plays no role in the following so we set the temperature as $k_{\rm B}T=1$ without any loss of generality.
The equilibrium distribution of the mixture is described by a probability density
\begin{equation} 
    p_{\TF}(\CC, \CF) = \frac{1}{\Xi_{\TF}}e^{\muB N + \muS n - U_{\TF}(\CC, \CF)}
    \label{equ:pf}
\end{equation}
where the subscript $\TF$ indicates that we refer to the FG model, $\muB,\muS$ are the chemical potentials for the large and small particles and $\Xi_{\TF}$ is the grand canonical partition function.
The particles are hard (non-overlapping) so the potential energy is 
\begin{equation}
    U_{\TF}(\CC, \CF) =
    \begin{cases}
        \infty, & \text{if any particles overlap,} \\
        0,      & \text{otherwise}.
    \end{cases}
\end{equation}
This $p_{\TF}$ is normalised as $1 = \int p_{\TF}(\CC, \CF) {\rm d}\CC {\rm d}\CF$, the precise meaning of these integrals is given in Appendix \ref{app:DetailsModel}.  

Within this setting, the dimensionless parameters of the model are the ratio of the particle diameters  $\sigB / \sigS$, the system size parameter $L/\sigB$, and the two chemical potentials $\muS,\muB$. 
In practice, $\muS$ is more naturally parametrised by the associated reservoir volume fraction $\etaS$, which we relate to $\muB$ via an accurate equation of state \cite{kolafa2004accurate}.

Our multi-level method is designed for accurate estimates of properties of the large particles.  Specifically, we consider observable quantities of interest
$A = A(\CC)$ that only depend on the large particles.  (Examples are discussed in the next Section, see also Fig.~\ref{fig:ExampleProperties}.)  Our aim is to compute the equilibrium average of $A$, that is
\beq
\langle A \rangle_{\TF}  =  \int A(\CC) p_{\TF}(\CC, \CF) {\rm d}\CC {\rm d}\CF.
\label{eqn:FineAverage}
\eeq
Since $A(\CC)$ does not depend on $\CF$, it is natural to define the marginal distribution for the big particles
\begin{equation}\label{eqn:marginal}
    p_{\TF}(\CC) = \int p_{\TF}(\CC, \CF) \id \CF,
\end{equation}
so that $\langle A \rangle_{\TF}  =  \int A(\CC) p_{\TF}(\CC) {\rm d}\CC$.  A similar situation occurs in the context of statistics, where one seeks to analyse the behaviour of a few quantities of interest in a high-dimensional system: in that context, the small-particle degrees of freedom in \eqref{eqn:FineAverage} would be referred to as \emph{nuisance parameters}.  This means that their values are not required to compute the quantity of interest, but their statistical properties strongly affect the average of this quantity.  

\subsection{Coase-grained model}

If samples for the marginal distribution $p_{\TF}(\CC)$ could be generated by an MC method for the big particles alone, this would make the system much more tractable by simulation.
This is a central idea in coarse-grained modelling\cite{noid2008multiscale}.
However, the complexity of packing of the small hard spheres means that $p_{\TF}(\CC)$ is a complex distribution, and it is not possible to sample it exactly.
A great deal of effort has gone into developing CG models that approximate this distribution with high accuracy\cite{dijkstra1999phase,roth2000depletion,ashton2011depletion}. 

A suitable CG model is an equilibrium distribution with probability density 
\begin{equation} \label{eqn:CGSystem}
    p_{\TC}(\CC) = \frac{1}{\Xi_{\TC}}e^{\muB N  - U_{\TC}(\CC)}
\end{equation}
where $\Xi_{\TC}$ is the partition function, and the CG (effective) interaction energy is 
\begin{equation}
\label{equ:UC}
    U_{\TC}(\CC) = N \Delta\mu + \sum_{i=1}^{N-1}\sum_{j=i+1}^{N} V_2(|\bfR_i - \bfR_j|),
\end{equation}
where $V_2$ is a pairwise interaction potential.  Averages with respect to the CG model are denoted as
\beq
\langle A \rangle_{\rm C} = \int A(\CC) p(\CC) \id\CC.
\eeq

For a suitably chosen $V_2$, the coarse distribution $p_{\TC}(\CC)$ can be an accurate approximation to $p_{\TF}(\CC)$.
For the CG model in this work, we take the accurate potential $V_2 = V_{\text{RED}}$, developed by Roth, Evans, and Dietrich \cite{roth2000depletion}.
Following Ref.~\onlinecite{kobayashi2021critical}, we choose $\Delta\mu$ such that the distributions of $N$ coincide for FG and CG models.

\subsection{Benchmark system: parameters and observables}\label{sec:example}

Throughout the paper, we benchmark our numerical methods by considering the hard-sphere mixture with fixed parameters, as follows.
We take the ratio of particle sizes $(\sigB/\sigS)=10$, the linear size of the periodic system is $L=31\sigS$, and the small-particle (reservoir) volume fraction is $\etaS=0.2$.  This volume fraction is large enough to generate a significant depletion attraction between the large particles, but not strong enough to cause de-mixing of the large and small particles\cite{kobayashi2021critical}.

Aspects of the CG and FG models are illustrated in Fig.~\ref{fig:OverviewMethod}(a-c), for these parameters.  In particular, we show representative configurations of the CG and FG models, as well as a plot of the RED potential.  While direct GCMC sampling of the full mixture is possible in principle, it should be apparent from Fig.~\ref{fig:OverviewMethod}(c) that this would be intractable, because insertion of large particles in such a fluid is hardly ever possible.  Advanced MC methods\cite{ashton2011grand,ashton2011depletion} might be applicable but these tend to struggle when the volume fraction gets large.  This motivates the development of two-level and multi-level methods.

\begin{figure}
    \centering
    \includegraphics[width=\columnwidth]{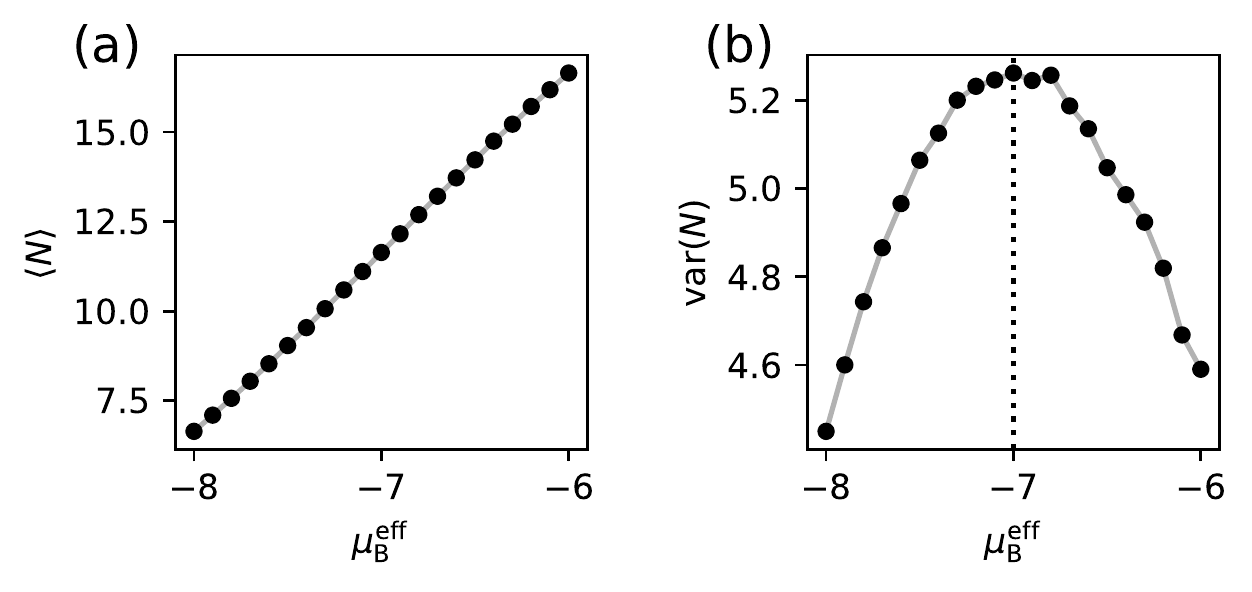}
    \caption{
        Properties of a big-particle-only hard sphere model with RED potential when varying the effective chemical potential $\mu_{B}^{\text{eff}}$ as defined in the main text.
        (a) The average number of big particles $\langle N\rangle $.
        (b) The variance of the number of big particles $\var(N)$, which is maximised around $\mu_B^{\text{eff}} = -7$.
    }
    \label{fig:ChoiceOfMu}
\end{figure}

Fig.~\ref{fig:ChoiceOfMu} highlights properties of the distribution of the number of big particles for the CG model when varying the effective large-particle chemical potential $\muB^{\text{eff}} = \muB - \Delta\mu$.
In particular, 
Fig.~\ref{fig:ChoiceOfMu}(b) shows that increasing $\muB^{\text{eff}}$ in the CG model leads to a non-monotonic behaviour in the variance of the particle number $N$ (analogous to the compressibility of the model).  This maximum indicates that the system has a tendency for de-mixing at larger $\etaS$ (one expects a divergent compressibility at the critical point, if one exists).  In the following, we fix $\muB$ at the value corresponding to this maximum -- the relatively large fluctuations at this point are challenging for the multi-level model, because the distributions $p_{\TC}(\CC)$ and $p_{\TF}(\CC)$ are broader, requiring good sampling.
The corresponding CG system has an average of $N \approx 11.6$ big particles, occupying around $20\%$ of the available volume.

\begin{figure}
    \centering
    \includegraphics[width=\columnwidth]{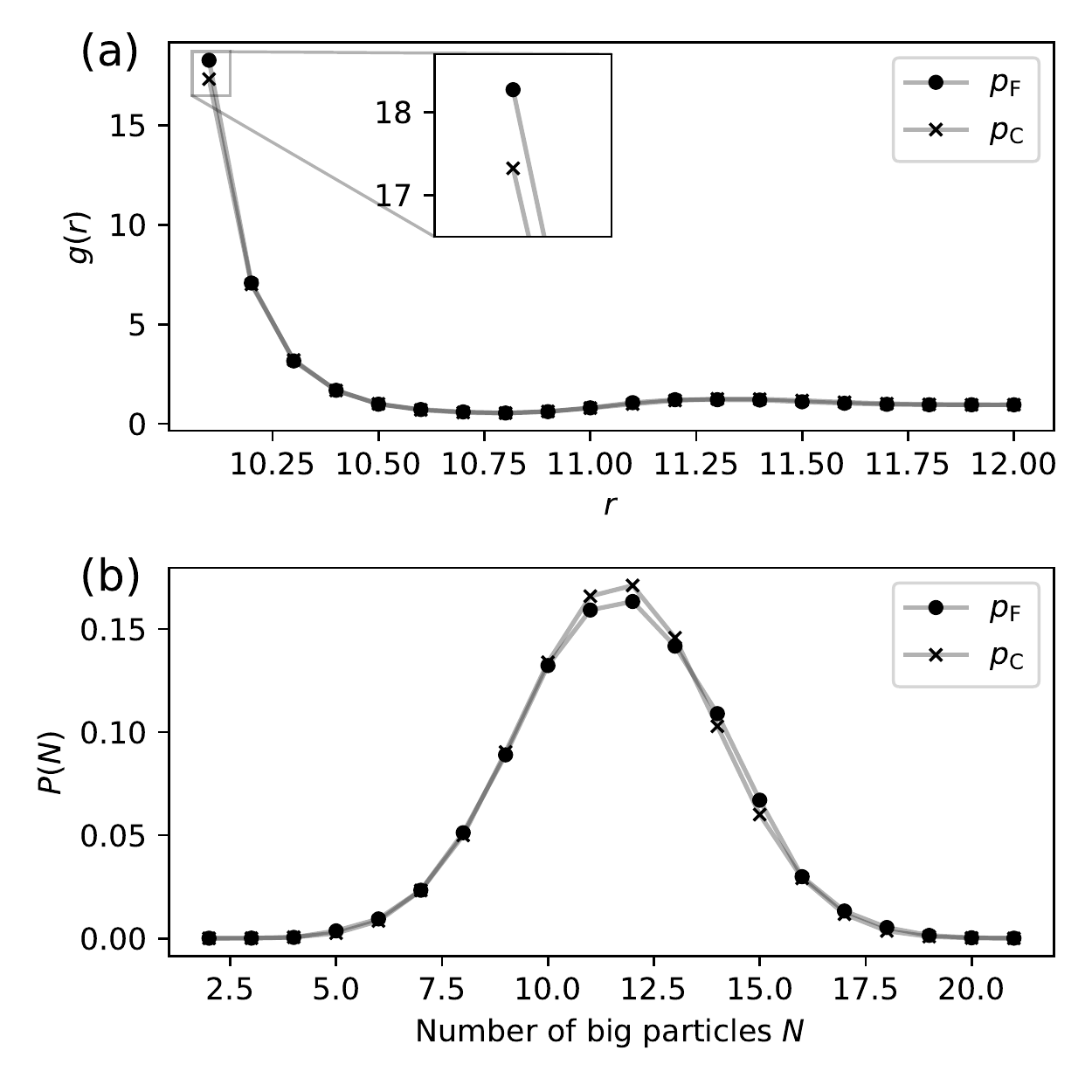}
    \caption{
        Two quantities of interest for the binary hard-sphere system, computed for the CG and FG model from Section \ref{sec:example}.
        (a) The big-particle pair correlation function $g(r)$. Apart from underestimating its value at the touch of two big particles, the CG approximation captures the behaviour accurately.
        (b) The distribution of the number of big particles $N$. By the choice of $\Delta \mu$, the average number of big particles of the CG and FG models coincide.
        Both models are clearly well below the critical point of demixing.
    }
    \label{fig:ExampleProperties}
\end{figure}

For the specific quantities that we will compute for this mixture, Fig.~\ref{fig:ExampleProperties} shows the expectations of the big-particle pair correlation function $g(r)$ and the distribution of the number of big particles $P(N)$.
Results are shown for both CG and FG models (in the FG case, results are computed using the two-level method).
For both quantities of interest, the CG model provides an accurate but not exact description of the model.
In particular, the CG model underestimates the pair correlation at the point where two big particles are in contact.
The distributions of the number of big particles in Figure \ref{fig:ExampleProperties}(b) are both unimodal: both the FG and CG systems are well below the critical point of demixing.

Compared to the critical hard-sphere mixture discussed in Ref.~\onlinecite{kobayashi2021critical}, the system we consider here is smaller and has a lower volume fraction $\etaS$ of the small particles.
This is still challenging for conventional Monte Carlo algorithms, but can be simulated fast enough to evaluate the performance and compare the computational methods discussed here.
Furthermore, the lower small-particle volume fraction helps with the construction of the intermediate level in Section \ref{sec:IntermediateLevel}, whose underlying approximation decays as $\etaS$ increases, see Appendix \ref{app:DetailsIntermediateLevel}.

\section{Multilevel simulation} \label{sec:method}

\subsection{Overview}

This section reviews the two-level method of Refs.~\onlinecite{kobayashi2019correction,kobayashi2021critical}, and then lays out its three-level extension.  The presentation of the method is intended to be generic and applicable to a variety of systems.  However we first introduce the key ideas using the example and illustrations of Fig.~\ref{fig:OverviewMethod}, for the hard-sphere mixture.

The two-level method is constructed with the scale separation of the mixture in mind: it splits the simulation of the big and small spheres into two stages by first simulating a CG system of large particles alone, and computing  $\langle A \rangle_{\TC}$.  
Then, differences between $\langle A \rangle_{\TC}$ and $\langle A \rangle_{\TF}$ are computed by a reweighting (importance sampling) method.
The weight factors for this computation are obtained by an annealing step, where the small particles are slowly inserted into the system, with the large particles held fixed (see Fig.~\ref{fig:OverviewMethod}(e)).
The advantage of this procedure is that large particle motion only happens in the CG simulation where the small particles are absent -- there is no scale separation in this case so simulations are tractable.  Similarly, insertion of the small particles happens in a background of fixed large particles, so these annealing simulations do not suffer long time scales associated with large-particle motion.  This makes for tractable simulations in scale-separated systems, as long as the CG model is sufficiently accurate: see Refs.~\onlinecite{kobayashi2019correction,kobayashi2021critical} for further discussion.

In practice, the simulation effort for two-level computations is dominated by the annealing step.  The weighting factors are required to high accuracy, which means that the annealing must be done gradually.  Moreover, the weights are subject to numerical uncertainties that tend to be large in systems with many small particles.  This limits the method to systems of moderate size, with moderate $\etaS$, see  Ref.~\onlinecite{kobayashi2021critical}.

We show in this work that such problems can be reduced by breaking the annealing process into several stages -- this is the idea of the three-level method (Fig.~\ref{fig:OverviewMethod}(e)).
Specifically, we start (as before) with a population of configurations of the CG model.  We perform a first annealing step where the small particles are added in regions that are close to large ones.
The information from this step is used in a resampling process, which partially corrects the coarse-graining error by discarding some of the configurations from the population, and duplicating others.
(This idea is similar to go-with-the-winners~\cite{grassberger2002go}.)
Finally, the second annealing step inserts the small particles in the remaining empty regions, arriving at configurations of the FG model.  Hence the end point is the same as the two-level method, but the annealing route is different.  

In practice the effectiveness of the three-level method relies on a clear physical understanding of the intermediate (partially-inserted) system, in order to decide which configurations to discard in the resampling step.  For the hard-sphere case, that issue will be discussed in Sec.~\ref{sec:IntermediateLevel}; a more general discussion is given in Sec.~\ref{sec:conclusions}.  The remainder of this Section describes the two- and three-level methods in more detail.

\subsection{Two-level method} \label{sec:TwoLeve}
\newcommand{\kk}{\kappa}
\newcommand{\Wn}{\hat{W}^{\rm n}}

We review the two-level method of Refs.~\onlinecite{kobayashi2019correction,kobayashi2021critical}.
For a general presentation, we assume that CG and FG models exist with configurations $\CC$ and $\XX=(\CC,\CF)$ respectively. 
In the case of hard spheres, $\CC$ and $\CF$ correspond to configurations of the large and small spheres respectively.

The two-level method is an importance sampling\cite{robert2004monte} (or reweighting) computation, closely related to the free-energy perturbation method of Zwanzig\cite{zwanzig1954high}.
We use the grand canonical Monte Carlo (GCMC) method to sample $M_{\TC}$ configurations from $p_{\rm C}$, these are denoted by $\CC^1,\CC^2,\dots,\CC^{M_{\TC}}$.
Then, the CG average can be estimated as
\begin{equation} \label{eqn:CGEstimator}
    \hat A_{\TC} = \frac{1}{M_{\TC}} \sum_{j=1}^{M_{\TC}} A(\CC^j). 
\end{equation}
As the sampling is increased ($M_{\TC} \to \infty$) we have $\hat A_{\TC} \to  \langle A \rangle_{\TC}$.
However, if the coarse-graining error
\begin{equation} \label{eqn:CGError}
    \Delta = \langle A \rangle_{\TF} - \langle A \rangle_{\TC}
\end{equation}
is significant then $\hat A_{\TC}$ does not provide an accurate estimate of $\langle A \rangle_{\rm F} $.

To address this problem, we use an annealing procedure based on Jarzynski's equality \cite{jarzynski1997nonequilibrium} that starts from a coarse configuration $\CC$ and populates the fine degrees of freedom $\hat \CF$; at the same time, it generates a random weight $\hat{W}(\CC)$ with the property that
\begin{equation} \label{eqn:unnormalisedWeight}
    \langle \hat W(\CC) \rangle_{\rm J}
        = \frac{\xi  p_{\rm F}(\CC)}{p_{\rm C}(\CC)}  ,
\end{equation}
where the angle brackets with subscript J indicate an averaging over the annealing process (analogous to Jarzynski's equality\cite{jarzynski1997nonequilibrium}), and $\xi$ is a constant (independent of $\CC$).
The details of the annealing process are given in Appendix~\ref{app:FreeEnergy}.
It is applied to a set of $M_{\TF}$ coarse configurations, again denoted by $\CC^1, \CC^2, \dots, \CC^{M_{\TF}}$, which are typically a subset of the $M_{\TC}$ CG configurations above.

For later convenience, we define
\beq
    \Wn(\CC) = \hat{W}(\CC)/\xi \; .
\eeq
In practical applications, the constant $\xi$ is not known but its effect can be controlled by defining the self-normalised weight
\begin{equation} \label{eqn:NormalisedWeight}
    \hat{w}(\CC^j) = \frac{\hat{W}(\CC^j)}{\frac{1}{M_{\TF}} \sum_{i=1}^{M_{\TF}} \hat{W}(\CC^i)}.
\end{equation}
Since the $\CC^j$ are representative of $p_{\rm C}$, the denominator in $\hat{w}$ converges to $\xi$ as $M_{\TF}\to\infty$ and so $\hat{w}(\CC^j) \to \Wn(\CC^j)$.
Then, the estimator
\begin{equation} \label{eqn:ISEstimator}
    \hat A_{\TF} = \frac{1}{M_{\TF}} \sum_{j=1}^{M_{\TF}} \hat{w}(\CC^j) A(\CC^j)
\end{equation}
converges to $\langle A \rangle_{\rm F}$ as $M_{\TF} \to \infty$.
(In the case that $\hat{W}$ is not random then this procedure recovers the free energy perturbation theory of Zwanzig\cite{zwanzig1954high}.)

The annealing process has one useful additional property: Let the joint probability density for the weight and the fine degrees of freedom be $\kk(\hat{W},\hat{\CF} | \CC)$, which is normalised as $\int \kk(\hat W, \hat \CF | \CC) \id \hat \CF \id \hat W = 1$.  
We show in Appendix~\ref{app:FreeEnergy} that
\beq
    \int \hat W \kk(\hat W, \hat \CF \mid \CC)  \id \hat W = \frac{   \xi p_{\TF}(\CC , \hat \CF) }{ p_{\TC}(\CC) }.
    \label{equ:jarz-fine-property}
\eeq
This formula is the essential property of the annealing procedure, which is required for the operation of the method.
Additionally integrating over $\hat \CF$ shows that \eqref{equ:jarz-fine-property} ensures that \eqref{eqn:unnormalisedWeight} also holds.
This means in turn that if $B = B(\CC, \CF)$ is an observable quantity that depends on both coarse and fine degrees of freedom then
\begin{equation}
  \hat{B}_{\rm F} = \frac{1}{M_{\TF}} \sum_{j=1}^{M_{\TF}} \hat w(\CC^j)B(\CC^j,\hat \CF^j). 
  \label{equ:hatB-2}
\end{equation}
converges to $\langle B \rangle_{\rm F}$ as $M_{\TF} \to \infty$.

This method can be easily improved without extra computational effort.  
The key idea~\cite{giles2008multilevel,hoang2013complexity,dodwell2015hierarchical} is to estimate the FG average as the sum of the CG average and the coarse-graining error \eqref{eqn:CGError}
\begin{equation}
    \langle A \rangle_{\TF} = \langle A \rangle_{\TC} + \Delta.
\end{equation}
Then use importance sampling to estimate $\Delta$, as
\begin{equation}\label{eqn:ErrorEstimator}
    \hat \Delta = \frac{1}{M_{\TF}} \sum_{j=1}^{M_{\TF}} \left( \hat w(\CC^j) - 1 \right) A(\CC^j).
\end{equation}
Finally, a suitable estimator for the FG average is obtained by combining the estimate of the coarse-graining error with the corresponding CG quantity:
\begin{equation}
    \hat A_{\TF, \Delta} = \hat A_{\TC} + \hat \Delta. 
    \label{equ:A-TML}
\end{equation}
This estimator converges to $\langle A \rangle_{\rm F}$ in the limit where $M_{\TC}, M_{\TF} \to \infty$. 
As discussed in Ref.~\onlinecite{kobayashi2019correction},
the variance of the estimate $\hat \Delta$ is typically smaller than that of $\hat A_{\TF}$, and the CG estimate $\hat A_{\TC}$ is cheap to compute accurately.
Thus, the combined difference estimator $\hat A_{\TF, \Delta}$ is typically more accurate at fixed computational cost.

The importance sampling methodology has a useful physical interpretation, which we explain for the example of the hard-sphere mixture.
If we consider a fixed configuration of the large particles, then the grand canonical partition function for the small particles is
\beq
    \Xi[\CC, \mu_S] = \int e^{\muS n - U_{\TF}(\CC, \CF)} \id \CF.
    \label{eqn:GCPotential}
\eeq
As the system is annealed (the small particles are inserted), we estimate \eqref{eqn:GCPotential} by a free-energy method based on Jarzynski's equality \cite{jarzynski1997nonequilibrium}, see Appendix~\ref{app:FreeEnergy} for details.
Since the annealing is stochastic, this yields an estimate of the partition function, which we denote by $\hat \Xi[\CC, \mu_S]$.
Moreover, this estimate is unbiased $\langle \hat \Xi[\CC, \mu_S] \rangle_{\rm J} = \Xi[\CC, \mu_S]$.
Hence we can take
\begin{equation} \label{eqn:defUnnormalisedWeight}
    \hat W(\CC) = \hat \Xi[\CC, \mu_{S}] e^{U_{\TC}(\CC)}
\end{equation}
and using  (\ref{eqn:marginal},\ref{eqn:CGSystem}), we see that \eqref{eqn:unnormalisedWeight} holds, with $\xi = (\Xi_{\rm C} / \Xi_{\rm F})$.

Physically, the CG model is constructed so that the Boltzmann factor $e^{-U_{\TC}(\CC)}$ is a good estimate of the small-particle partition function $\Xi[\CC, \mu_{S}]$.  If this is the case then the model is accurate.  The two-level methodology uses estimates of the small-particle partition function (or, equivalently, their free energy) and compares it with the assumptions that were made about this quantity in the CG model.  By analysing the differences between these quantities, the differences between CG and FG models can be quantified.
The effectiveness of this method for numerical simulation of the mixtures of large and small particles was discussed in Refs.~\onlinecite{kobayashi2019correction,kobayashi2021critical}.

\begin{figure}
    \centering
    \includegraphics[width=0.9\columnwidth]{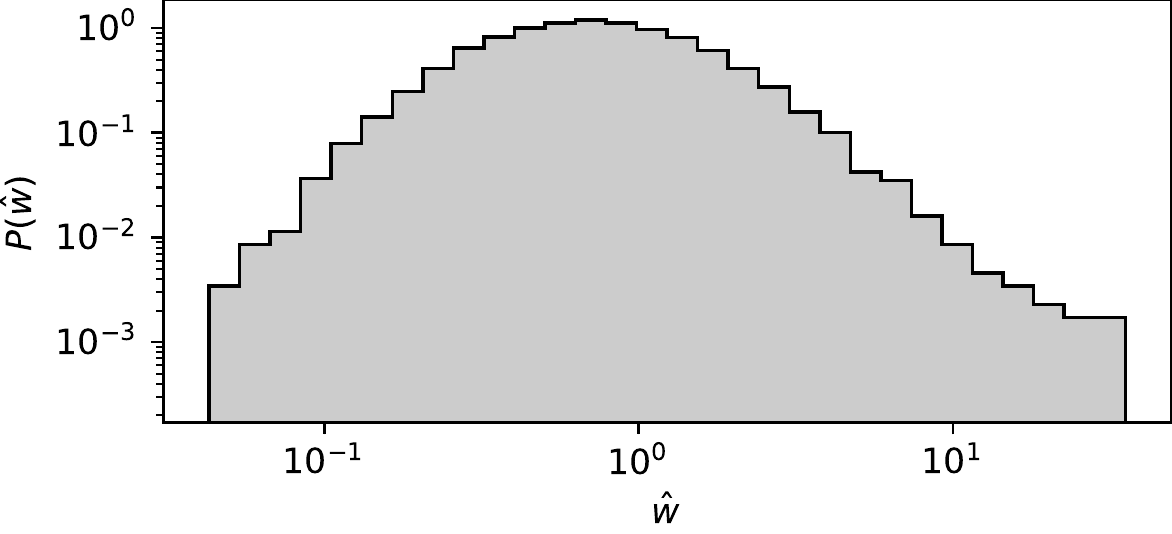}
    \caption{
        The empirical distribution of weights $\hat w(\CC)$ of the two-level method for $18000$ coarse samples $\CC \sim p_{\TC}$ applied to the example system from Section \ref{sec:example}.
    }
    \label{fig:Weights}
\end{figure}

The distribution of the importance weights $\hat w(\CC)$ impacts the accuracy of the resulting FG estimate $\hat A_{\TF}$.
Additionally, it serves as a useful indicator of the accuracy of the CG model and the variance of the free energy computation.
To give an example, we apply the two-level method to the example problem from Section~\ref{sec:example}.
In Figure \ref{fig:Weights}, we show the empirical distribution of $18000$ weights of the example system which are computed using an accurate annealing process; we use these computations as the reference solution in Section \ref{sec:NumericalResults}.
This illustrates a situation where the two-level method is applicable, where no single sample dominates and only very few samples have a weight larger than $10$.

If one considers less accurate CG models, the variance of the weights increases, and the tail of their distribution gets heavier.   Eventually, one would reach a situation where a few samples dominate the weighted sum \eqref{eqn:ISEstimator}.
For accurately computed weights $\hat w(\CC)$, such a breakdown of the two-level method indicates that the CG model is not sufficiently accurate.
This behaviour provides a useful feedback loop which can be used to iterate on the CG model itself \cite{kobayashi2019correction}.

\subsection{Three-level method}\label{sec:threelevel}

We now present the three-level method for estimation of $\langle A(\CC) \rangle_{\TF}$.

\subsubsection{Coarse level} \label{par:CoarseLevel}

We start by generating $M_{0}$ samples of the CG model, denoted by
$    \CC^1_0, \dots, \CC_0^{M_{0}} $ 
The subscript indicates the step within the algorithm (which is $0$ for the initial sampling of coarse configurations).
The CG average of $A$ can be estimated similarly to (\ref{eqn:CGEstimator}):
\begin{equation}
    \hat A_{\TC}^{\text{3L}} = \frac{1}{M_{0}} \sum_{j=1}^{M_{0}} A(\CC^j_0) \; .
\end{equation}

\subsubsection{Intermediate level}
\label{sec:3-int}

\begin{figure}
    \centering
    \includegraphics[width=\columnwidth]{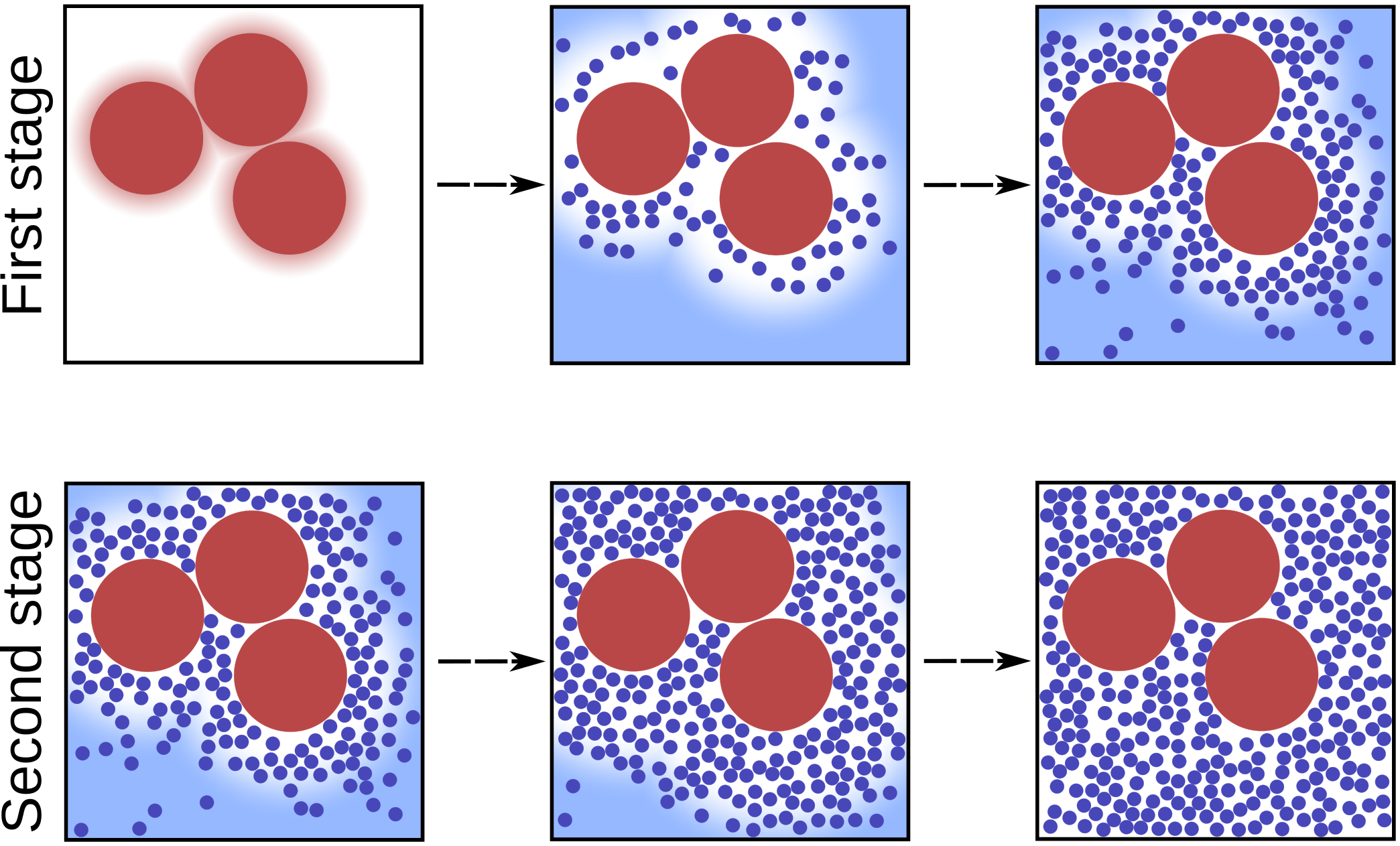}
    \caption{Schematic representation of the small particle insertion process during the two stages  of the three level method.
    }
    \label{fig:AISStages}
\end{figure}

In addition to the CG and FG models, the three-level method also relies on an intermediate set of configurations, which correspond
 in the hard-sphere mixture to the system where the small particles have been inserted in regions close to the large ones, see Fig.~\ref{fig:AISStages}.
This state is described by an equilibrium probability distribution
\begin{equation}\label{eqn:AbstractIntermediateDistribution}
    p_{\TI}(\CC, \CF) = \frac{1}{\Xi_{\TI}} e^{\muB N + \muS n - U_{\TI}(\CC, \CF)}.
\end{equation}
where $U_{\TI}(\CC, \CF)$ is an interaction energy.
Its construction for the hard-sphere mixture will be discussed in Sec.~\ref{sec:IntermediateLevel}, below.  

The first annealing step of the three-level algorithm applies the two-level method, with the FG distribution $p_{\rm F}$ replaced by $p_{\rm I}$.  
This part of the algorithm closely follows the previous section, we give a brief discussion which mostly serves to fix notation.
We start with a set of $M_{1}$ coarse configurations which are samples of $p_{\rm C}$; they are denoted by
 $   \CC^1_1, \CC^2_1, \dots, \CC_1^{M_{1}}$ 
where now the subscript $1$ indicates the intermediate stage of the three-level method.
(These will typically be a subset of the configurations that were generated on the coarse level.)

For each coarse configuration $\CC_1$, we anneal the fine degrees of freedom of the system $\hat \CF_1$ to arrive at the intermediate level and generate a random weight $\hat W_1(\CC_1)$ with the property
\begin{equation}
    \langle \hat W_1(\CC_1) \rangle_{\rm J} = \frac{\xi_1 p_{\TI}(\CC_1)}{p_{\TC}(\CC_1)},
    \label{eqn:AvgW1}
\end{equation}
with a constant $\xi_1$ independent of $\CC_1$.
(For the hard spheres, we recall that particles are inserted preferentially in regions close to large ones, this is illustrated in the top row of Figure \ref{fig:AISStages}.)

As before, we define
$
    \hat W^{\rm n}_1(\CC_1) = \hat W_1(\CC_1) / \xi_1.
$
Again the constant $\xi_1$ is generally not known, so we define the self-normalised weight
\begin{equation}
    \hat w_1(\CC_1^j) = \frac{\hat W_1(\CC_1^j)}{\frac{1}{M_{1}} \sum_{i=1}^{M_1} \hat W_1(\CC_1^i)},
\end{equation}
which converges to $\hat W^n_1(\CC_1^j)$ as $M_{1} \to \infty$.
Then, the estimator
\begin{equation}
    \hat A^{\text{3L}}_{\TI} = \frac{1}{M_1} \sum_{j=1}^{M_1} \hat w_1(\CC_1^j) A(\CC_1^j)
\end{equation}
converges to $\langle A \rangle_{\TI}$ as $M_1 \to \infty$.
Similar to (\ref{equ:jarz-fine-property}), the joint probability density $\kk_1(\hat W_1, \hat \CF_1 \mid \CC_1)$ of the weight and fine degrees of freedom at the intermediate level, defined by the annealing process, fulfils
\beq
    \int  \hat W_1 \kk_{1}(\hat W_1, \hat \CF_1 \mid \CC_1) \id \hat W_1 = \frac{ \xi_1 p_{\TI}(\CC_1, \hat \CF_1) }{ p_{\TC}(\CC_1) }
    \label{equ:jarz-fine-property-int}.
\eeq
Hence, similar to \eqref{equ:hatB-2} we also obtain
\begin{equation}
  \hat{B}^{\rm 3L}_{\TI} = \frac{1}{M_1} \sum_{j=1}^{M_1} \hat w_1(\CC_1^j)B(\CC_1^j, \hat \CF_1^j) 
\end{equation}
which converges to $\langle B \rangle_{\TI}$ as $M_1 \to \infty$.

\subsubsection{Fine level}
\label{sec:3-fine}

\begin{figure}
    \centering
    \includegraphics[width=\columnwidth]{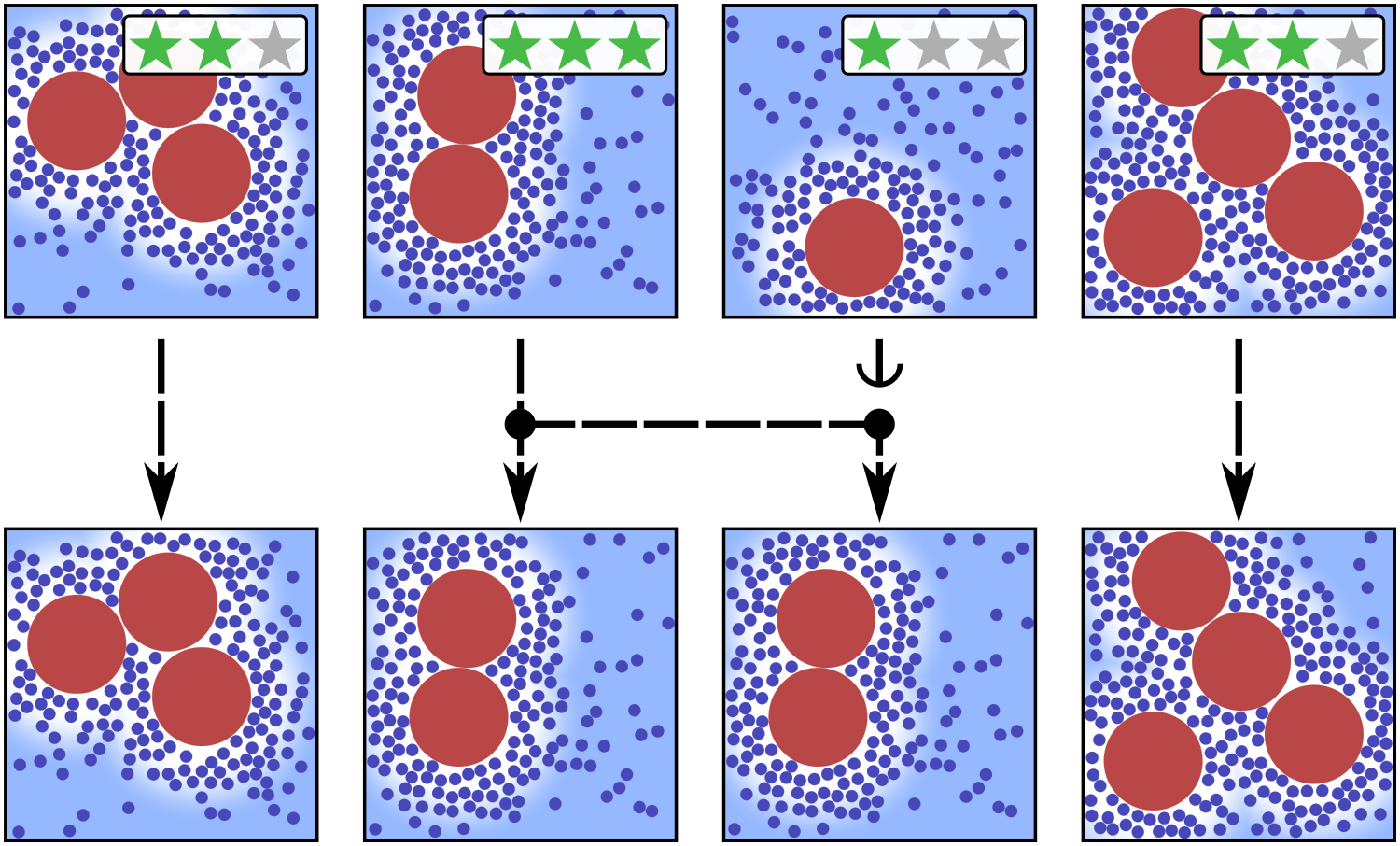}
    \caption{
        Visualisation of the resampling step.
        We start with a population of weighted configurations (top row), where the weighting is depicted by a star rating.
        The goal of the resampling step is to randomly transform the weighted population into an unweighted one that has, on average, the same empirical distribution.
        We achieve this by duplicating large-weight configurations and deleting small-weight configurations, yielding an unweighted population of configurations (bottom row).
    }
    \label{fig:Resampling}
\end{figure}

At the end of the intermediate level, we have $M_1$ large-particle configurations.
For each configuration $\CC_1^j$, the process of annealing to the intermediate level also provided the weight $\hat w_{1}(\CC_1^j)$ and the small-particle configuration $\hat \CF_1^j$.
This information can be used to build a set of configurations that are representative of $p_{\rm I}$.
This procedure is called \emph{resampling}, its validity in this example relies on the property \eqref{equ:jarz-fine-property-int} of the annealing procedure.
This is the part of the method that is similar to population-based sampling approaches such as SMC~\cite{hsu2011review} or go-with-the-winners~\cite{grassberger2002go}. 
The idea is that one should focus the effort of the annealing process onto coarse configurations which are typical of the full system, and to discard those which are atypical, see Fig.~\ref{fig:Resampling} for a visualisation of this step.

We write $\hat \XX_1^j=(\CC_1^j,\hat \CF_1^j)$ for the full configuration that is obtained by the annealing procedure at the intermediate level.
The resampled configurations will be denoted by $\XX_2^1,\XX_2^2,\dots,\XX_2^{M_2}$; they are representative of the intermediate level $p_{\TI}$.
There are $M_2$ of them, and the subscript $2$ indicates the final stage of the three-level method.
The simplest resampling method (\emph{multinomial resampling}) is that each $\XX^i_2$ is obtained by copying one of the $\hat \XX_1^j$, chosen at random with probability $\hat{w}_{1}(\CC^j_1)$.
In applications, one typically replaces this by a lower variance resampling scheme like residual resampling, see Ref.~\onlinecite{douc2005comparison} for a comparison of commonly used variants.

We then perform the second annealing step that starts from an intermediate level configuration $\XX_2 = (\CC_2, \CF_2)$ and anneals the fine degrees of freedom from the intermediate to the fine level, yielding $\hat \CF_2$ and a weight $\hat W_2(\XX_2)$, details are given in Appendix \ref{app:FreeEnergy}.
For the hard sphere system, this involves further insertion of small particles, to fill the system and generate realistic configurations of the full mixture.
This procedure is shown in the bottom row of Figure \ref{fig:AISStages}.

Since the starting point of the annealing procedure is $\XX_2$, the joint probability density of the annealing process $\kk_2(\hat W_2, \hat \CF_2 \mid \XX_2)$ depends on both large and small particles.
Therefore, the analogue of (\ref{equ:jarz-fine-property-int}) requires an additional average over the small particles of the starting configuration:
\begin{multline}
    \int \hat W_2 \kk_2(\hat W_2, \hat \CF_2 \mid \CC_2, \CF_2)  p_{\TI}(\CF_2 \mid \CC_2)  \id \hat W_2 \id \CF_2 
     \\
    \quad = \frac{\xi_2 p_{\rm F}(\CC_2, \hat \CF_2) }{ p_{\rm I}(\CC_2)} 
    \label{equ:WF-ave}
\end{multline}
for some constant $\xi_2$.
Note that $p_{\rm I}(\CF | \CC) = p_{\rm I}(\CC,\CF)/p_{\rm I}(\CC)$.
Similar to \eqref{eqn:AvgW1}, the weights $\hat W_2(\XX_2)$ have the property
\begin{equation}
    \int \langle \hat W_2(\CC_2, \CF_2) \rangle_{\rm J} p_{\TI}(\CF_2 \mid \CC_2) \id \CF_2
        = \frac{\xi_2 p_{\TF}(\CC_2)}{p_{\TI}(\CC_2)}.
        \label{eqn:MarginalIntegralW2}
\end{equation}
From here, we proceed as before.
We define the normalised weight
$    \Wn_2(\XX_2) = \hat W_2(\XX_2) / \xi_2
$
and its self-normalised estimate
\begin{equation}
    \hat w_{2}(\XX_2^j) = \frac{\hat W_2(\XX^j_2)}{\frac{1}{M_2} \sum_{i=1}^{M_2} \hat W_2(\XX^i_2)}.
\end{equation}
Since the $\XX_2^j$ are representative of $p_{\rm I}$, it follows from \eqref{equ:WF-ave} that observables of the coarse system $A$ can be estimated as
\begin{equation} \label{eqn:IntermediateEstimate}
    \hat A_{\TF}^{\text{3L}} = \frac{1}{M_2} \sum_{j=1}^{M_2} \hat{w}_2(\XX_2^j) A(\CC_2^j),
\end{equation}
which converges to $\langle A \rangle_{\TF}$ as $M_2 \to \infty$.
Similar to \eqref{equ:hatB-2}, we can also obtain a consistent FG estimates of observable quantities $B$ that depend both on coarse and fine degrees of freedom by
\begin{equation}
    \hat B_{\TF}^{\text{3L}} = \frac{1}{M_2} \sum_{j=1}^{M_2} \hat w_2(\XX_2^j) B(\CC_2^j, \hat \CF_2^j).
    \label{eqn:IntermediateEstimateB}
\end{equation}

Following the same variance reduction strategy as in Section \ref{sec:TwoLeve}, we can define a difference estimator of the FG average, which is expected to have lower statistical uncertainty: let
\begin{align}
    \hat \Delta_{\TI}^{\text{3L}} &= \frac{1}{M_{1}} \sum_{j=1}^{M_{1}} \left( \hat w_1(\CC_1^j) - 1 \right) A(\CC_1^j), \label{eqn:3LDiffEst1}\\
    \hat \Delta_{\TF}^{\text{3L}} &= \frac{1}{M_{2}} \sum_{j=1}^{M_{2}} \left( \hat w_{2}(\XX^j_2) - 1 \right) A(\CC_2^j). \label{eqn:3LDiffEst2}
\end{align}
Then
\begin{equation} \label{eqn:Multilevel3LEstimator}
    \hat A_{\TF, \Delta}^{\text{3L}} = \hat A_{\TC}^{\text{3L}} 
            + \hat \Delta_{\TI}^{\text{3L}} 
            + \hat \Delta_{\TF}^{\text{3L}}
\end{equation}
is a consistent estimator of $\langle A\rangle_{\rm F}$, analogous to (\ref{equ:A-TML}).

\subsubsection{General features of the three-level method} \label{sec:properties}

A few comments on the three-level method are in order.
First, there is a simple generalisation to four or more levels by splitting the annealing procedure into more than two stages. 
As such, the method is an example of a sequential Monte Carlo (SMC) algorithm (which is sometimes more descriptively referred to as sequential importance sampling and resampling~ \cite{cappe2005inference,del2006sequential,ionides2006inference,hsu2011review}).
We note from (\ref{eqn:unnormalisedWeight}) that the weights obtained from the annealing step are random, this is not the standard situation in SMC but similar ideas have been previously studied in Refs.~\onlinecite{fearnhead2008particle,fearnhead2010random,naesseth2015nested,rohrbach2022convergence}.
Combining an SMC algorithm with a difference estimate as in \eqref{eqn:Multilevel3LEstimator} has been investigated in Refs.~\onlinecite{jasra2017multilevel,beskos2017multilevel,del2017multilevel}.

Second, we observe that the key distinction between the two- and three-level algorithms is the resampling step at the intermediate level.
Without this, the three-level method reduces to a simple two-level method with an arbitrary stop in the middle of the annealing process.
As noted above, the resampling process is designed to partially correct differences between the CG and FG models.
This relies on a good accuracy of the intermediate level (otherwise the wrong configurations might be discarded, which hinders numerical accuracy).
On the other hand, we note that for sufficiently large numbers of samples $M_{0},M_{1},M_{2}$, the method does provide accurate FG estimates, even if the CG and intermediate level models are not extremely accurate.
The distinction between the different methods comes through the number of samples that are required to obtain accurate FG results.

Third, note that the ideal situation for difference estimation is that the three terms in (\ref{eqn:Multilevel3LEstimator}) get successively smaller.
That is, the coarse estimate is already close to $\langle A\rangle_{\TF}$, the intermediate-level estimate provides a large part of the correction, and the fine-level correction is small. 
In this case, it is natural to use a tapering strategy where the number of samples used at each level decreases
\begin{equation}
    M_{0} > M_{1} > M_{2}.
\end{equation}
This allows a fixed computational budget to be distributed evenly between the various levels, to minimise the total error.

\section{Construction of the intermediate level} \label{sec:IntermediateLevel}

As noted above, the intermediate probability distribution $p_{\rm I}$ must be designed carefully, in order for the resampling part of the three-level method to be effective.
We now describe how this is achieved for the hard sphere mixture.

To motivate the intermediate level, recall Fig.~\ref{fig:AISStages}, and note that defining a suitable CG model is equivalent to an estimate of the small-particle free energy in the final (fully inserted) state.  The physical idea of the intermediate level is that the free energy associated with the first stage of insertion may be hard to estimate (because of the complicated packing of the small particles around the large ones), but the free energy difference associated with the second stage should be easier (because it corresponds to insertion into large empty regions where the packing of the small particles is similar to that of a homogeneous fluid, whose free energy can be estimated based on analytic approximations).
A combination of these ideas yields an intermediate level that represents the big particle statistics more accurately than the CG model.
Similar ideas have been considered before in multi-scale simulation \cite{rafii1998multi,praprotnik2005adaptive,praprotnik2007macromolecule}, in particular the problem of estimating the small-particle free energy has some similarities to estimation of solvation free energies (where the depletant here plays the role of a solvent).

We start by analysing the small particles, so we fix the large particles in some configuration $\CC$.
The idea of the intermediate level is to first insert small particles only in a region close to the large particles $\CC$, and then use this information to make the intermediate marginal distribution $p_{\TI}(\CC)$ match the FG marginal $p_{\TF}(\CC)$ as closely as possible.
The structure of the intermediate level is depicted in the bottom row of Fig.~\ref{fig:OverviewMethod}(e), and an example configuration is shown in Fig.~\ref{fig:OverviewMethod}(d).
We implement this idea by introducing an effective (one-body) potential that acts on the small particles.
We first define
\begin{equation}
    \text{dist}(\bfr, \CC) = \min_{j=1, \dots, N} |\bfr - \bfR_j|
\end{equation}
to be the distance from the point $\bfr$ to the nearest large particle.
Small-particle insertion is suppressed in regions far from large particles by a potential energy term
 \begin{equation}
 \label{eqn:tildeU}
    \tilde U(\CC, \CF) = \sum_{j=1}^{n} E_{\CC}(\bfr) 
\end{equation}
where 
\begin{equation} \label{eqn:SmallParticlePotential}
    E_{\CC}(\bfr) = \eps(\text{dist}(\bfr, \CC)) 
\end{equation}
and the function
\begin{equation} \label{eqn:CosinePotential}
    \eps(r) 
    = 
    \begin{cases}
        0, & r < \delta_{\text{free}}, \\
       s  \sin^2 \left[ \frac{(r - \delta_{\text{free}})\pi}{2l} \right] , 
            & \delta_{\text{free}} \leq r < \delta_{\text{free}}+l, \\
        s, & r \geq \delta_{\text{free}}+l
    \end{cases}
\end{equation}
interpolates from zero (for small distances $r$) to the value $s$ at large $r$.
This function acts as a smoothed out step function, where $\delta_{\rm free}$ is the position of the step and $l$ its width.
In Figs.~\ref{fig:OverviewMethod}(e) and \ref{fig:AISStages}, areas where $E_{\CC}(\bfr) > 0$ are indicated by blued shaded regions, in which the insertion of small particles is suppressed.

Then define a grand-canonical probability distribution for the small particles in the partially-inserted (intermediate) system as
\begin{equation} \label{eqn:IntermediateEnsemble}
     \tilde p_{\TI}(\CF \mid \CC ) = \frac{1}{\tilde \Xi_{\TI}[\CC,\muS]} e^{\muS n - U_{\TF}(\CC, \CF) - \tilde U(\CC, \CF)}.
\end{equation}
This distribution is normalised as $\int  \tilde p_{\TI}(\CF \mid \CC ) \id \CF=1$.  It depends on the three parameters $s,\delta_{\rm free},l$, as well as the underlying parameters of the hard sphere mixture model.  

The next step is to construct the weights $\hat{W}_1(\CC)$.
For consistency with (\ref{eqn:AbstractIntermediateDistribution}), we write the intermediate-level distribution in the form
\begin{equation} \label{equ:int-cor}
      p_{\TI}(\CC,\CF) = \frac{1}{\Xi_{\rm I}} e^{ \muB N + \muS n - U_{\TF}(\CC, \CF) - \tilde U(\CC, \CF) - \Phi^{\rm corr}(\CC) }.
\end{equation}
As discussed above, the term $\Phi^{\rm corr}$ should be designed so that the respective coarse-particle marginals  $p_{\rm I}(\CC)$ and $p_{\TF}(\CC)$ match as closely as possible.
Using (\ref{eqn:marginal},\ref{eqn:GCPotential},\ref{eqn:IntermediateEnsemble}), we can show that a perfect match requires $\Phi^{\rm corr}(\CC) = \Phi^{\rm ex}(\CC)$ with
\beq
  \Phi^{\rm ex}(\CC) = \log \frac{ \tilde \Xi_{\TI}[\CC,\muS] }{ \Xi_{\TF}[\CC,\muS] } - \phi_0,
  \label{equ:Phiex}
\eeq
where $\phi_0$ is an irrelevant constant.
Since the $\Xi$s in \eqref{equ:Phiex} are partition functions, determination of $\Phi^{\rm ex}$ reduces to computation of the free energy difference between the non-homogeneous small particle distributions of the partially- and fully-inserted system.
We now explain how $\Phi^{\rm corr}$ is defined, as an approximation to $\Phi^{\rm ex}$.

\subsection{Square-gradient approximation of a non-homogeneous hard sphere fluid} \label{sec:SGApprox}
\label{sec:sg}

\newcommand{\Cz}{{\bf 0}}
\newcommand{\press}{\mathfrak{p}}
\newcommand{\gam}{\mathfrak{g}}
\newcommand{\Egen}{\mathcal{E}}

As a preliminary step for estimating $\Phi^{\rm ex}$, we first consider the grand potential $\Phi$ for the small particles, in a system with no large particles, where the small particles feel an (arbitrary) smooth potential $\Egen=\Egen(\bfr)$.
The grand potential of this system is 
\begin{equation}
    \Phi[\Egen; \muS] = - \log \int e^{\muS n - U_{\TF}(\Cz, \CF)
                     - \sum_{j=1}^{n} \Egen(\bfr_j)} \id \CF.
     \label{eqn:PerturbedSmallSphereFluid}
\end{equation}
where $\Cz$ indicates the large-particle configuration with no particles at all ($N=0$).

If $\Egen$ varies slowly in space, a simple approach to this integral is to assume that the system is locally the same as a homogeneous system in equilibrium -- similar to the local density approximation~\cite{evans1979nature}.
In this case
\begin{equation}\label{eqn:eqintegral}
  \Phi[\Egen; \muS]
        \approx -\int_V \mathfrak{p}(\muS - \Egen(\bfr)) \id \bfr
\end{equation}
where $\press$ is the pressure, expressed as a function of the chemical potential.  

However, this approximation is not sufficiently accurate for the current application.  
To this end, we include a correction to account for inhomogeneities, as a squared gradient term:
$\Phi[\Egen; \muS]  \approx \Phi^{\rm sq}[\Egen; \muS] $ with
\begin{equation}\label{eqn:sgapprox}
  \Phi^{\rm sq}[\Egen; \muS] =  
       - \int_V \mathfrak p(\muS - \Egen(\bfr)) 
        + \gam(\muS - \Egen(\bfr)) |\nabla \Egen(\bfr)|^2 \id \bfr.
\end{equation}
(Within a gradient expansion, this is the first correction that is consistent with rotational and inversion symmetry.)

We show in Appendix \ref{app:DetailsApprox}, that $\gam$ can be estimated as
\begin{equation}
    \gam(\mu) = \frac{3\eta}{2\pi \sigma_{S}^3}  \frac{\partial^2}{\partial q^2} S(\mu; q) \Big\rvert_{q=0},
    \label{equ:gam}
\end{equation}
where $S(\mu;q)$ is the structure factor of the small hard-sphere system.
For a numerical estimate of this $\Phi^{\rm sq}$, we estimate the pressure $\press$ by the accurate equation of state from Ref.~\onlinecite{kolafa2004accurate}, and $\gam$ is estimated from (\ref{equ:gam}) using the structure factor from Ref.~\onlinecite{de2004structure}.
A numerical example demonstrating the accuracy of this second order approximation for a non-homogeneous hard-sphere fluid can be found in Appendix \ref{app:exampleSG}.

\subsection{Definition of $\Phi^{\rm corr}$}

\begin{figure}
    \centering
    \includegraphics[width=0.95\columnwidth]{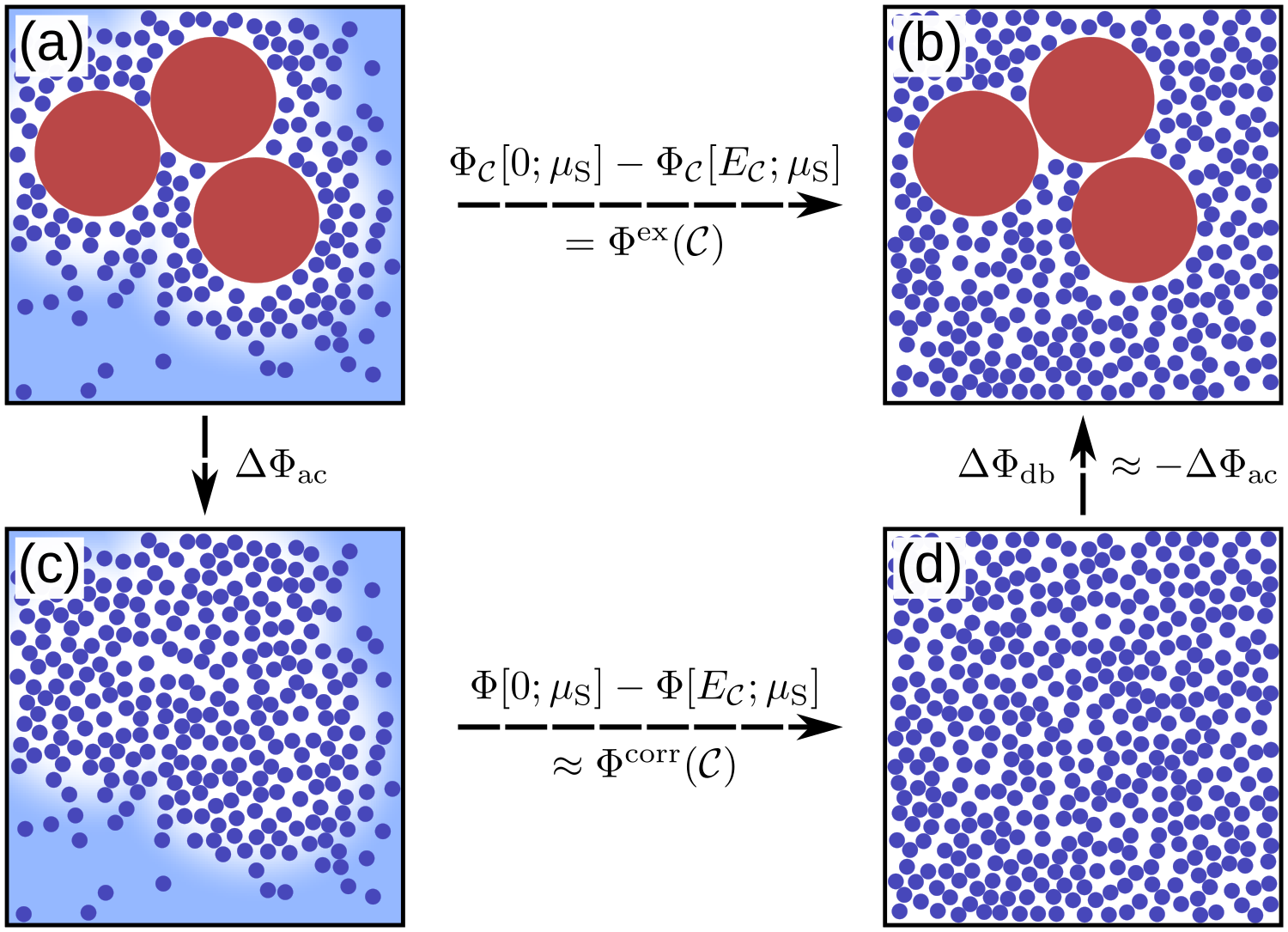}
    \caption{
        Illustration of the computation of $\Phi^{\rm corr}$, which is an estimate of
        the free energy difference $\Phi^{\rm ex}$  between panels (a,b), see \eqref{eqn:DefDeltaPhi}.
        As described in the text, this difference is computed as a sum of three differences, using the integration path (a,c,d,b).
        The free energy difference between (c,d) is $\Phi[ 0 ; \muS ] - \Phi[ E_\CC ; \muS ]$.  We make the approximation that the differences between (a,c) and (d,b) are equal and opposite, this should be accurate if the shaded blue regions  in panel (a) are well-separated in space from the large particles.  Combining this assumption with the square gradient approximation (\ref{eqn:sgapprox}) yields $ \Phi^{\rm corr}(\CC)$ in (\ref{eqn:PhiCorr1}) as a numerically tractable estimate of $\Phi^{\rm ex}$.
    }
    \label{fig:FreeEnergyApproximation}
\end{figure}

We are now in a position to approximate $\Phi^{\rm ex}$ in terms of $\Phi^{\rm sq}$. 
This (analytical) calculation is illustrated in Fig.~\ref{fig:FreeEnergyApproximation}.
We require an estimate of $\Phi^{\rm ex}$, which is the free-energy difference between the partially-inserted and fully-inserted systems in panels (a,b).
This is achieved as a sum of three free-energy differences.
In the first step, the large particles are removed and the small-particle fluid is re-equilibrated, to fill up the remaining space, leading to panel (c).
Then, the confining potential $\tilde U$ is removed and the small particles fully inserted, leading to (d).
Finally, the large particles are re-inserted and the small particles re-equilibrated again, leading to (b).  

To make this precise, define $\Phi_\CC[ \Egen ; \muS ]$ as the grand potential of the small particles in the potential $\Egen$, where the large particles are also included, with configuration $\CC$.  Then the desired free energy difference between panels (a,b) is
\beq \label{eqn:DefDeltaPhi}
    \Phi^{\rm ex}(\CC) = \Phi_\CC[ 0 ; \muS ] - \Phi_\CC[ E_\CC ; \muS ] 
\eeq
where we took $\phi_0=0$.

From the definitions in Sec.~\ref{sec:sg}, the free energy difference between panels (c,d) is $\Phi[ 0 ; \muS ] - \Phi[ E_\CC ; \muS ]$, from (\ref{eqn:tildeU},\ref{eqn:PerturbedSmallSphereFluid}).
Our central approximation is that the free energy difference between panels (a,c) is (approximately) equal and opposite to the difference between (d,b), because the local environment of the large particles is the same in both cases.
(The only differences are in regions far from any large particles.)  
At this level of approximation, the free energy differences between (a,b) and (c,d) are equal:
\beq \label{eqn:ApproxOfFEWithC}
   \Phi^{\rm ex}(\CC) \approx \Phi[ 0 ; \muS ] - \Phi[ E_\CC ; \muS ] \; . 
\eeq
Finally, the right hand side can be estimated by the square gradient approximation (\ref{eqn:sgapprox}), yielding
$\Phi^{\rm ex}(\CC) \approx   \Phi^{\rm corr}(\CC) $ with
\beq  \label{eqn:PhiCorr1}
    \Phi^{\rm corr}(\CC) = \Phi^{\rm sq}[ 0 ; \muS ] - \Phi^{\rm sq}[ E_\CC ; \muS ] \; .
\eeq

Operation of the three-level method requires numerical estimates of this $\Phi^{\rm corr}$, which includes the integral  in (\ref{eqn:sgapprox}).  Moreover, its value is exponentiated when computing weight factors $\hat{W}$, so these numerical estimates are required to high accuracy.  This is a non-trivial requirement because the integrand is constant on regions far from the big particles, but it varies much more rapidly when these particles are approached.  In such situations, adaptive quadrature schemes are appropriate: we use the \texttt{cuhre} algorithm of the \texttt{cuba} library\cite{hahn2005cuba} which uses globally adaptive subdivision to refine its approximations in the relevant regions of space.
%
Note however that while the choice of the numerical integrator influences the intermediate level, small errors in estimation of this integral will be corrected by the second annealing step, so such errors do not affect the consistency of our numerical estimators.

Given this choice of $\Phi^{\rm corr}(\CC) $, the intermediate level distribution $p_{\rm I}$ of (\ref{equ:int-cor}) has been completely defined, although it still depends on the three parameters $\delta_{\rm free},s,l$ that appear in the function $\eps(r)$.
We also note that given the approximations made, it is not expected that this $p_{\rm I}$ is optimal (its marginal $p_{\TI}(\CC)$ does not match $p_{\TF}(\CC)$ perfectly).
The next subsection discusses the parameter choices, and some possibilities for correction factors that can be added to $\Phi^{\rm corr}$, in order to address specific sources of error.

\subsection{Variants of the intermediate-level distribution}

\begin{figure}
    \centering
    \includegraphics[width=\columnwidth]{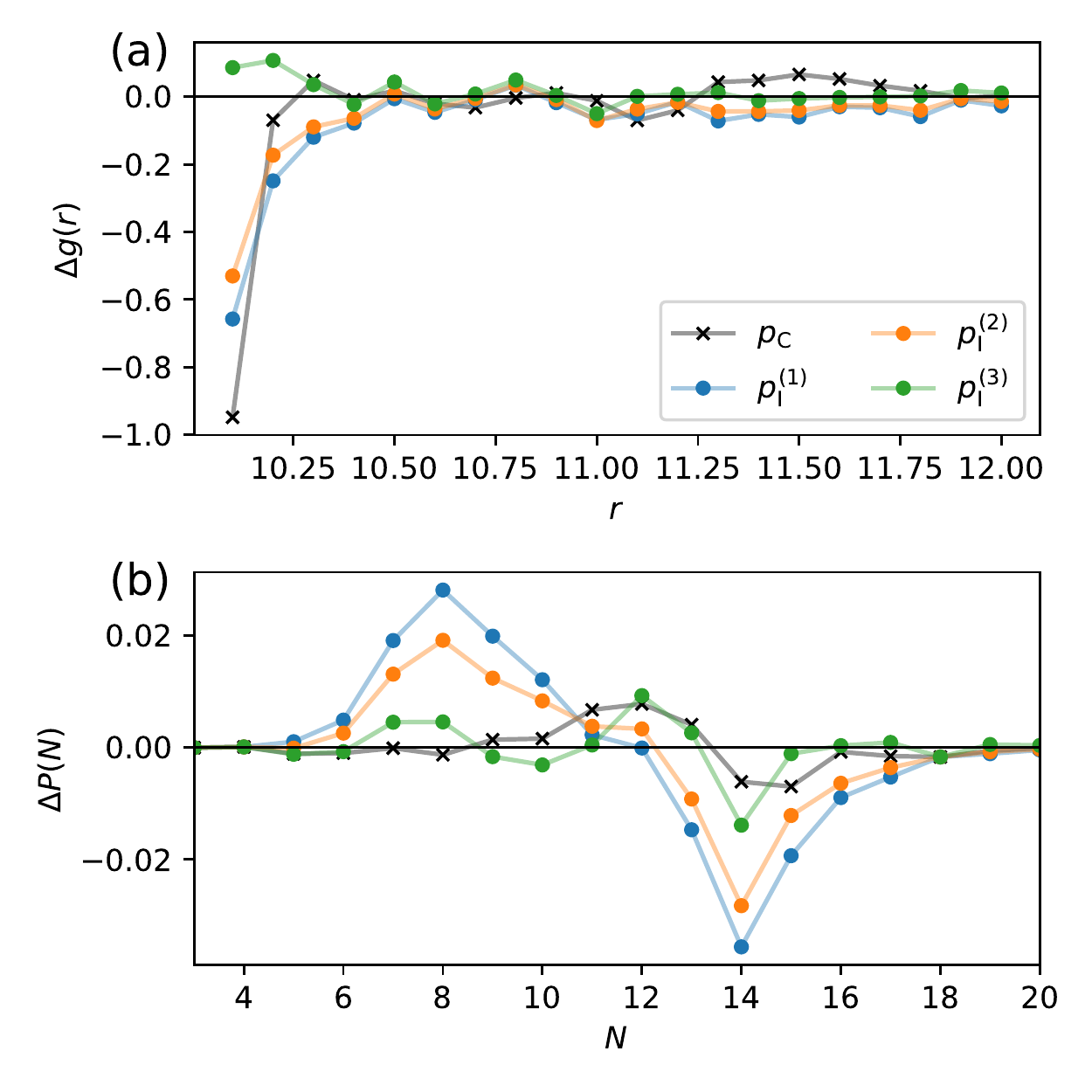}
    \caption{
        Estimated accuracy of the CG and the intermediate levels from the main text for the example from Section \ref{sec:example}.
        We show the difference between the respective CG and intermediate level estimates and the true FG estimate for the pair correlation function $g(r)$ in (a) and the distribution of big particles in (b).
        The FG estimates of these quantities of interest are shown in Figure \ref{fig:ExampleProperties}.
    }
    \label{fig:AccuracyIntermediateLevel}
\end{figure}

In fixing the parameters $\delta_{\rm free},s,l$, several considerations are relevant.  First, if $s$ is too small or $\delta_{\rm free}$ is too large, the potential $E_\CC$ has little effect on the system and the small particles are not restricted to be close to the large ones.  In this case $p_{\rm I}$ ends up close to $p_{\rm F}$ and there is little benefit from the intermediate level.  On the other hand, the accuracy of $\Phi^{\rm sq}$ is greatest when the gradient of the potential $E_\CC$ is small, this favours small $s$ and large $l,\delta_{\rm free}$.  In practice, it is also convenient if the two annealing stages insert similar numbers of particles, so that their computational costs are similar.
For the example system of Section \ref{sec:example}, we will present results for a suitable parameter set
\begin{equation}\label{eqn:ILParameters}
    \delta_{\text{free}} = 0.5\sigS, \;\;\; s = 4.4, \;\;\; l = 3.5\sigS.
\end{equation}
We have also tested other values, a few comments are given below.

We will consider several variants of the intermediate level.  We denote by $p_{\rm I}^{(1)}$ the distribution defined by (\ref{equ:int-cor},\ref{eqn:PhiCorr1}), with parameters (\ref{eqn:ILParameters}).
Fig.~\ref{fig:AccuracyIntermediateLevel} shows how the quantities of interest differ between the CG and FG models, and the corresponding differences between the intermediate level and the FG model.
Here $\Delta g(r)$ is the difference between $g(r)$ for the FG model and the distribution of interest (which is either the CG distribution $p_{\rm C}$ or one of the variants of the intermediate distribution).
And $\Delta P(N)$ is the corresponding difference in the probability that the system has $N$ large particles.  

For the value of $g(r)$ at contact, we see that the intermediate level $p_{\rm I}^{(1)}$ corrects around half of the deviation between CG and FG models.
However, the probability distribution of the $N$ has the opposite situation, that the intermediate level is \emph{less} accurate than the  CG model.  [This is partly attributable to the fact that $\Delta\mu$ in Eq.~(\ref{equ:UC}) has been chosen to make the CG model accurate.]

\begin{figure}
    \centering
    \includegraphics[width=0.8\columnwidth]{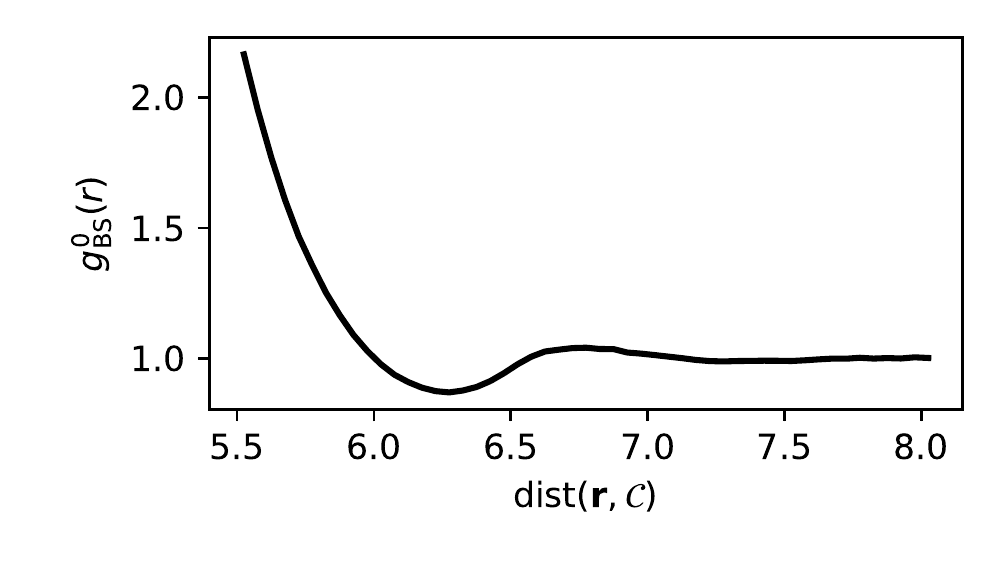}
    \caption{The layering of small particles ($\sigma_S = 1$) around one big particle ($\sigma_B = 10$) at volume fraction $\eta_{S} = 0.2$.
    The displayed pair correlation function $g_{\rm BS}(r)$ given the radius from the centre of the big particle shows that small particles form layers of higher and lower concentration that vanish with increased distance from the big particle.}
    \label{fig:Layering}
\end{figure}

To explore the behaviour of the intermediate level, we constructed two variants of $p_{\rm I}$.  The aim is to understand why $ p_{\rm I}^{(1)}$ has inaccuracies, and to (partially) correct for them.
There are two main approximations in the intermediate level $p_{\rm I}^{(1)}$: the first is (\ref{eqn:ApproxOfFEWithC}) and the second is that $\Phi[E_{\CC},\muS]$ can be approximated by the square-gradient approximation (\ref{eqn:sgapprox}).
The first approximation neglects a significant physical phenomenon in these systems, which is a layering effect of the small particles around the large ones.
This is illustrated in Fig.~\ref{fig:Layering} by the radial distribution function $g_{\rm BS}^0$ between large and small particles (measured in a system with a single large particle).
One sees that there is typically an excess of small particles close to the large ones, followed by a deficit ($g_{\rm BS}^0(r)<1$), and a (weak) second layer.   

For (\ref{eqn:ApproxOfFEWithC}) to be accurate, the intermediate level should have enough small particles to capture this layering, so that the particles being inserted in the second annealing stage are not strongly affected by the presence of the large particles.
However, computational efficiency requires that $\delta_{\rm free}$ is not too large, so these layers are not fully resolved at the intermediate level.
To partially account for this effect, we make an ad hoc replacement of $\muS$ in \eqref{eqn:sgapprox} by an effective chemical potential $\mu_{\rm lay}(\textbf{r})$, which is chosen such that the corresponding reservoir volume fraction $\eta^{\rm lay}_{\rm S}(\textbf{r})$ satisfies
\beq
    \frac{ \eta^{\rm lay}_{\rm S}(\textbf{r}) }{ \etaS }  
        = g_{\rm BS}^0( \text{dist}(\bfr, \CC)).
\eeq
In estimating the free energy of the small particles that are inserted in the second level of annealing, this adjustment to $\Phi^{\rm sq}$ helps to counteract the error made in (\ref{eqn:ApproxOfFEWithC}), leading to an updated potential $\Phi^{\rm corr, 2}$.
The intermediate level constructed in this way is denoted by $p_{\rm I}^{(2)}$.
The results of Fig.~\ref{fig:AccuracyIntermediateLevel} show that this variant is (somewhat) more accurate than $p_{\rm I}^{(1)}$.
However, the intermediate level still tends to have a smaller number of large particles than the full (FG) mixture.

\begin{figure}
    \centering
    \includegraphics[width=\columnwidth]{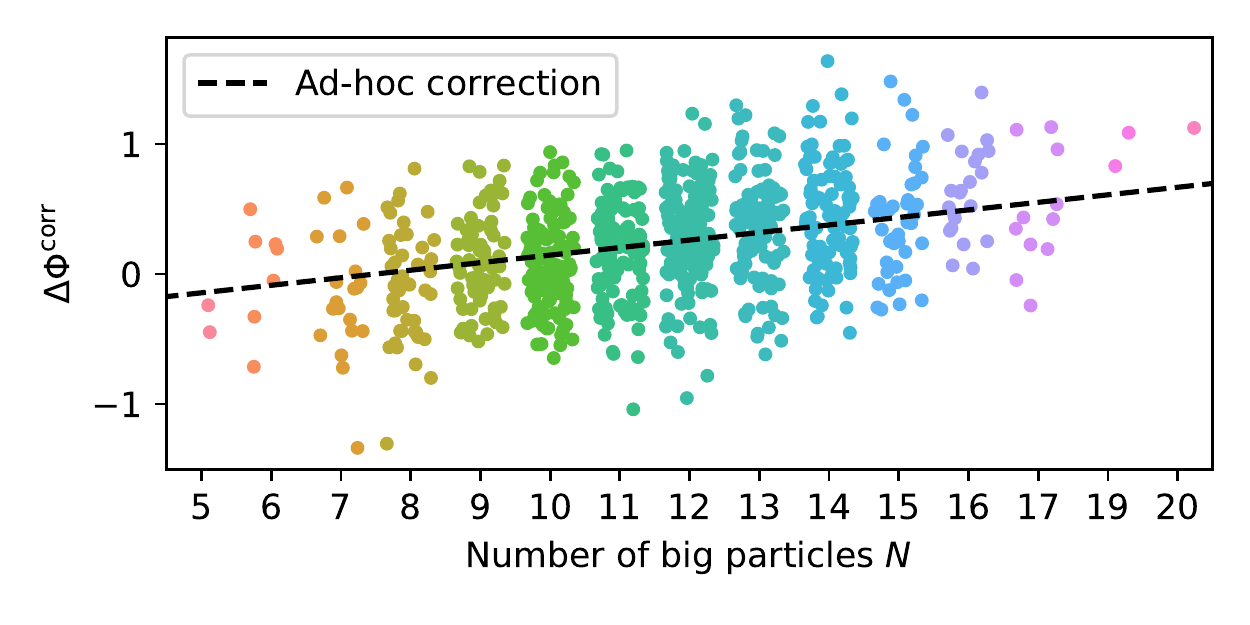}
    \caption{The error of the free energy prediction for $800$ configurations sampled from the coarse distribution $p_{\TC}$, grouped by their number of big particles.
    Each dot represents the difference of an estimate of the predicted small-particle free energy used in the intermediate level $p_{\TI}^{(2)}$ against an estimate of the full free energy.
    We correct for the noticeable trend in the error with a linear ad-hoc correction term, displayed in black.}
    \label{fig:AdHocCorrection}
\end{figure}

To investigate this further, we took 800 representative CG configurations.
For each one, we estimate the error associated with the approximation (\ref{eqn:PhiCorr1})
\begin{equation}
    \Delta\Phi^{\rm corr} = \Phi^{\rm ex} - \Phi^{\rm corr, 2}.
\end{equation}
Results are shown in Fig.~\ref{fig:AdHocCorrection}.
One sees that the errors are of order unity (note that $\Phi^{\rm ex}$ itself is of order $10^4$ so this is a small relative error, see below); there is a systematic trend, that $\Phi^{\rm corr, 2}$ underestimates $\Phi^{\rm ex}$ when $N$ is large. 
To correct this error we introduce an additional correction term to $\Phi^{\rm corr,2}$
\begin{equation}
    \Phi^{\rm corr,3}(\CC) = \Phi^{\rm corr,2}(\CC) + \alpha_{\rm corr} N
\end{equation}
and denote the intermediate level constructed in this way by $p_{\rm I}^{(3)}$.

A least squares fit to Fig.~\ref{fig:AdHocCorrection} suggests to take $\alpha_{\rm corr}=0.076$; in practice this tends to over-correct the error in $\Phi^{\rm corr,2}(\CC)$ and we find better performance with a smaller value
\begin{equation}\label{eqn:ILParameters2}
    \alpha_{\text{corr}} = 0.058.
\end{equation}
However, the performance of the method depends only weakly on the specific choice of
$\alpha_{\text{corr}}$, this is discussed in  Appendix \ref{app:AdHocCorrection}.
For all following results, we define the intermediate level $p_{\TI} = p_{\TI}^{(3)}$ to use the potential $\Phi^{\rm{corr,3}}$.

\subsection{Discussion of intermediate level}

An important aspect of the three-level method is the self-consistency of the general approach.
The intermediate level variants $p_{\rm I}^{(1)}$ and $p_{\rm I}^{(2)}$ were constructed on a purely theoretical basis.
The corresponding results in Fig.~\ref{fig:AccuracyIntermediateLevel} indicated good performance, but that the distribution of $N$ had a systematic error.
This error was quantified precisely in Fig.~\ref{fig:AdHocCorrection}, which enabled an improvement to the intermediate level.
In principle, this procedure could be repeated to develop increasingly accurate variants of $p_{\rm I}$. 
That approach would be useful if (for example) one wanted to consider increasingly large systems, where the requirements for the accuracy of $p_{\rm I}$ become increasingly demanding.

One way to see the effect of system size is to note that Fig.~\ref{fig:AdHocCorrection} required the estimation of $\Phi_{\CC}[E_\CC,\muS]$ and $\Phi_{\CC}[0,\muS]$, whose values are of order \num{1e4}.
Since the free energies are exponentiated in the weights for resampling, an absolute error of $\pm1$ is required on these free energies, while their absolute values are extensive in the system size.
Hence one sees that accurate estimates of the free energy are required: their relative error is required to be of the order of the inverse volume of the system.

\section{Numerical tests}\label{sec:NumericalResults}

In this Section, we apply the three-level method to the example from Section \ref{sec:example} using the intermediate level from Section \ref{sec:IntermediateLevel}, with the parameters defined in \eqref{eqn:ILParameters} and \eqref{eqn:ILParameters2}.
The parameters and the annealing schedules are chosen such that, on average, the first and second step have the same computational effort, see Appendix \ref{app:FreeEnergy} for details.

It can be proven~\cite{rohrbach2022convergence} that the three-level method provides accurate results, in the limit where the population sizes $ M_{0}, M_1, M_2$ are all large.
In particular, we expect the estimators $\hat{A}^{\rm 3L}_{\TF}, \hat{A}^{\rm 3L}_{\TF,\Delta}$ to all obey central limit theorems (CLTs), the two-level estimators $\hat{A}_{\TF}, \hat{A}_{\TF,\Delta}$ behave similarly.
Detailed results are given in Sec.~\ref{sec:ConvergenceResults}.  
The important fact is that for large populations, the variances of the estimators behave as
\beq
    \Var(\hat{A}) \approx \frac{1}{M} \Sigma
    \label{equ:clt-gen}
\eeq
where $M$ is the relevant population size and $\Sigma$ is called the asymptotic variance (it depends on the observable $A$ and on which specific estimator is used).
In general, the estimators may have a bias, which is also of order $1/M$.
This means that the uncertainty in our numerical computations is dominated by the random error, whose typical size is $\sqrt{\Sigma/M}$, and the mean squared error is given by the variance $\text{MSE}(\hat A) = \Var(\hat A)$, to leading order.

This gives us an easy way to measure and compare the performance of the different estimators.  Suppose that we require an estimate of $A$ with a prescribed mean squared error.  The associated computational cost can be identified with the population size $M$, and is given by (\ref{equ:clt-gen}) as
$M \approx \Sigma/\text{MSE}(\hat A)$.  Clearly, estimators with small $\Sigma$ should be preferred.  In practice, we do not compare computational costs at fixed error, instead we compare variances $\Var(\hat{A})$ at fixed $M$.  For any two algorithms (and assuming that $M$ is large), the ratio of these variances approximates the ratio of the $\Sigma$'s, which can then be interpreted as a ratio of computational costs (at fixed MSE).  Numerical results are presented in Sec.~\ref{sec:variance}, below.

We note that the theoretical results for convergence do not require that the coarse or intermediate levels are accurate.
However, one easily sees~\cite{kobayashi2019correction} that serious inaccuracies in these levels lead to very large $\Sigma$.
In such cases, one may require prohibitively large populations to obtain accurate results.  

In this section, we demonstrate (for the example of Sec.~\ref{sec:example}) that we do not require very large populations for the three-level method, and that the numerical results are consistent with (\ref{equ:clt-gen}).
After that, we estimate the asymptotic variances for the two-level and three-level methods.
We will find that introducing the third level improves the numerical performance, corresponding to a reduction in $\Sigma$.

To this end, we investigate the pair correlation $g(r)$ of the big particles. 
As seen in Figure \ref{fig:AccuracyIntermediateLevel}(a), the coarse approximation of $g(r)$ has a substantial error, especially when two big particles are in contact.
To quantify this specific effect, we define the coordination number $N_c$, which is the number of large particles within a distance $r_1$ of a given large particle.
(For a given configuration, this quantity is estimated as an average over the large particles.  We take $r_1 \approx 10.73\sigS $ to be the first minimum of $g(r)$ of the CG model.)
For our example, the coordination number for the FG and CG systems are given by 
\begin{equation}
    \langle N_c \rangle_{\TF} \approx 1.61, \qquad\langle N_c \rangle_{\TC} \approx 1.56.
\end{equation}

\subsection{Accuracy of method} \label{sec:convergence}

\begin{figure}
    \centering
    \includegraphics[width=\columnwidth]{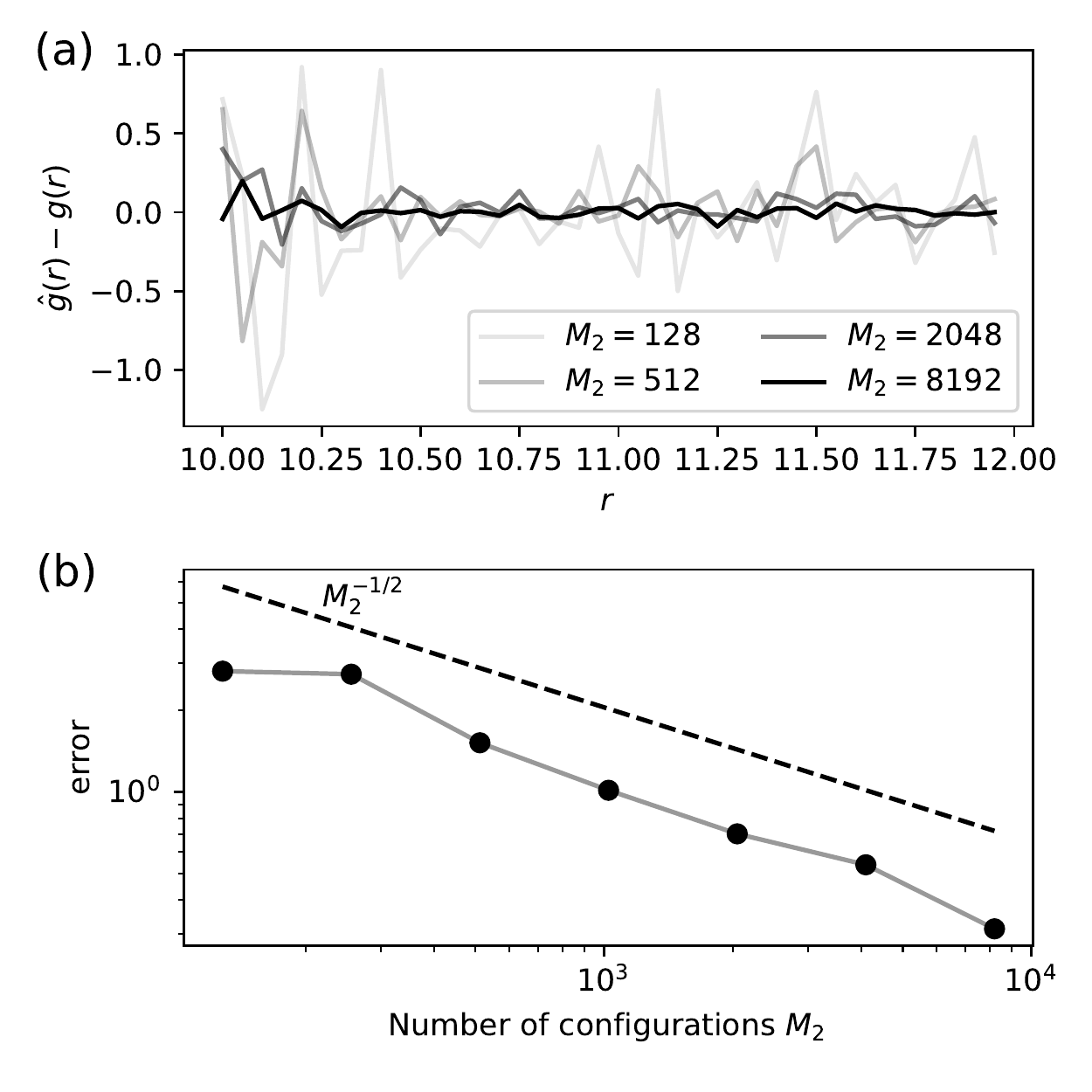}
    \caption{
        Estimating the pair-correlation function $g(r)$ from Figure \ref{fig:ExampleProperties}(a) with the two- and three-level method.
        (a) The difference of three-level estimates $\hat g(r)$ with increasing numbers $M_{1} = M_{2}$ number of particles against a reference value of $g(r)$ which was computed with the two-level method.
        (b) The error of the binned values in (a) as defined in \eqref{eqn:BinningError}.
        The dotted black line displays the expected asymptotic Monte Carlo convergence rate of $M_{2}^{-1/2}$.
    }
    \label{fig:Convergence}
\end{figure}

To illustrate the reliable performance of the method, we take a simple example with $M_0 = \num{4e5}$ and $M_{1} = M_{2}$ (no tapering) and we focus on the difference estimator $\hat A_{\TF, \Delta}^{\text{3L}}$, which we expect to be the most accurate.
The corresponding numerical estimate of $g(r)$ is denoted by $\hat g(r)$, binned using $40$ equidistant bins at positions $r_{j}$ between $r=10$ and $r=12$.
Figure \ref{fig:Convergence}(a) shows estimates of the difference between $\hat g(r)$ and its true value, as the population size increases.
(The FG result was estimated independently by the two-level method, using a large value of $M_{\TF}=\num{18000}$.)
A population $M_{2}$ of several thousand is sufficient for an accuracy better than $0.5$ in each bin of $g(r)$. 

For smaller $M_{2}$, fluctuations in the measured $\hat g(r)$ are apparent in Figure \ref{fig:Convergence}(a).
To estimate their size, we define the error for a single run of the three-level method by summing over the bins:
\beq
    ({\rm error})^2 = \sum_{j=1}^{40} |\hat g(r_j) - g(r_j)|^2.
    \label{eqn:BinningError}
\eeq
Hence, one expects from (\ref{equ:clt-gen}) that this error decays with increasing population, proportional to $M_{\TF}^{-1/2}$.
Figure \ref{fig:Convergence}(b) shows an estimate of \eqref{eqn:BinningError}, which is consistent with this expected scaling.

\subsection{Measurements of variances $\Sigma$} \label{sec:variance}

We now investigate whether the three-level method does indeed improve on the performance of the (simpler) two-level method of Refs~\onlinecite{kobayashi2019correction,kobayashi2021critical}.
The key question is whether the resampling step is effective in focussing the computational effort on the most important configurations of the big particles.

We recall from above that removing the resampling step from the three-level method leads to a two-level method, where the annealing process is paused at the intermediate level, and then restarted again.
In order to test the effect of resampling, we compare these two schemes, keeping the other properties of the algorithm constant, including the annealing schedule.
(To test the overall performance, one might also optimise separately the annealing schedules for the two-level and three-level algorithms, and compare the total computational time for the two methods to obtain a result of fixed accuracy.
However, such an optimisation would be very challenging, so instead we focus on the role of resampling.)

As a very simple quantity of interest, we take the co-ordination number $N_c$.
We run the whole algorithm $N_{\rm runs}$ independent times and we estimate $N_c$ for each run.
This can be done using several different estimates of $N_c$.
These are: (i) the two-level estimates $\hat{A}_{\rm F}$ and $\hat{A}_{\TF,\Delta}$ from (\ref{eqn:ISEstimator},\ref{equ:A-TML}); (ii) the corresponding three-level estimates $\hat{A}^{\rm 3L}_{\rm F}$ and $\hat{A}^{\rm 3L}_{\TF,\Delta}$ of (\ref{eqn:IntermediateEstimate},\ref{eqn:Multilevel3LEstimator}), in which we also vary the ratio $M_{1}:M_{2}$, to see the effects of tapering.

All comparisons are done with a fixed total computational budget.
We have chosen parameters such that the first and second annealing stage have the same (average) computational cost.
This means we need to hold $M_{\rm T} = (M_{1}+M_{2})/2$ constant during tapering.
The two-level method takes $M_{\TF} = M_{\rm T}$ (because the single step of annealing in the two-level method has the same cost as the two annealing steps of the three-level method).
For the coarse level estimates ${\hat A}_{\TC}$ and ${\hat A}_{\TC}^{\rm 3L}$ (which are used in computation of $\hat{A}_{\TF,\Delta}$ and $\hat{A}^{\rm 3L}_{\TF,\Delta}$), the CG computations are cheap so we take $M_0 = M_{\TC}=\num{6e6}$.
This is large enough that the numerical errors on these coarse estimates are negligible in comparison to the errors from higher levels.

\begin{figure}
    \centering
    \includegraphics[width=\columnwidth]{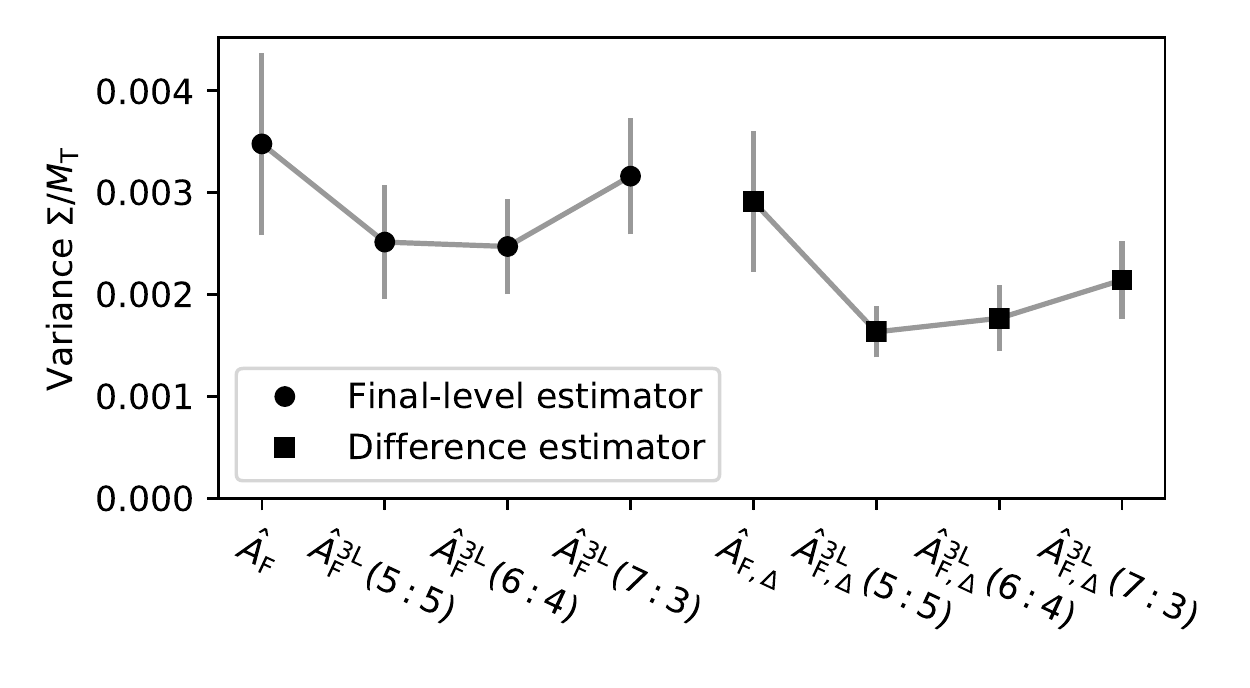}
    \caption{
        The sample variance of $N_{\rm runs} = 60$ independent estimates of the coordination number $N_c$.
        We compare results using a two-level method as well as three-level methods, with and without tapering.
        Further, we give the results for the final-level (left) and difference (right) variants of the estimator.
        The error bars are computed via bootstrap; their interpretation is however not obvious as the different estimators are highly correlated, see the main text.
     }
    \label{fig:ResultsStdDev}
\end{figure}

For each version of the algorithm, we measure the sample variance of the $N_{\rm runs}$ estimates.  
Results are shown in Figure \ref{fig:ResultsStdDev} for $N_{\rm runs}=60$ and $M_{\rm T} = 500$.
The error bars are computed by the bootstrap method\cite{efron1981nonparametric}.  
It is useful that the variance of all these estimators are expected to be proportional  $1/M_{\rm T}$: this means that reducing the variance by a factor of $\alpha$ requires that the computational effort is increased by the same factor.
Hence the ratio of variances of two estimators is a suitable estimate for the ratio of their computational costs.

When carrying out these runs, each estimator was computed by performing annealing on the same set of coarse configurations, to ensure direct comparability. 
(More precisely: we take a set of $700$ representative configurations which are used for the method with $M_{1}:M_{2}=7:3$,  other versions of the method used a subset of these $700$.)
In addition, it is possible to share some of the annealing runs when computing the different estimators (while always keeping the 60 different runs completely independent).
This freedom was exploited as far as possible, which reduces the total computational effort.  However, it does mean that the calculations of the different estimators are not at all independent of each other.

\subsection{Performance: discussion} \label{sec:variance-discuss}

All three-level estimators have a reduced standard deviation compared to their two-level equivalents, demonstrating the usefulness of the intermediate resampling step.
In all cases, the difference estimate outperforms its equivalent final-level estimate; this effect is stronger for the three-level estimate, providing evidence that the intermediate stop additionally improves the quality of the control variate in the difference estimates.

The effect of introducing tapering from $M_{\TI}=600$ to $M_{\TF}=400$ is difficult to assess, given the statistical uncertainties in this example.
The variance of the tapered final-level estimator is very close to the non-tapered one, despite averaging over fewer configurations.
This is possible since we start with more samples in the CG model which improves the sampling at the intermediate step, where we then resample to keep relevant configurations.
As the results for the $700$ to $300$ tapering shows, the tapering rate needs to be chosen carefully as a too aggressive rate can decrease the performance quickly.

Overall, the numerical tests in this section provide strong evidence of the benefit of the intermediate resampling.
For our example, switching from a two-level to a three-level difference estimator substantially reduces the variance, from around $0.0029$ for the two-level method to $0.0016$ at a fixed computational budget. As discussed just below \eqref{equ:clt-gen}, the ratio of these numbers can be interpreted as the ratio of costs for the two- and three-level method: the conclusion for this case is that including the intermediate level reduces the cost by approximately 45\%.  This demonstrates a significant speedup in this specific case, which provides a proof-of-principle of the approach.

\section{Convergence of the multilevel method} \label{sec:ConvergenceResults}

In Section \ref{sec:NumericalResults}, we have seen that the three-level method outperforms the two-level method in numerical tests, both for the final-level as well as the difference version of the estimator.
In this section, we provide convergence results for both algorithms, and compare their asymptotic performance as the number of configurations goes to infinity.

The proof is general, but it does require some assumptions on the models of interest.
First, for every allowed CG configuration (that is, configurations $\CC$ with $p_{\rm C}(\CC)>0$), we assume that the quantity of interest $A$ is bounded.
Also, the probability density $p_{\CC}(\CC)$ must be non-zero whenever $p_{\TI}(\CC)$ is non-zero, and similarly $p_{\TI}(\CC, \CF)$ must be non-zero whenever $p_{\TF}(\CC, \CF)$ is non-zero.

\subsection{Two-level method}

\newcommand{\dA}{A^{\rm r}}

The two-level method has been previously analysed in Ref.~\onlinecite{kobayashi2019correction}.
We summarise its key properties.
It was noted in Sec.~\ref{sec:TwoLeve} that $\hat{A}_{\rm F}\to \langle A\rangle_{\rm F}$ as $M_{\TF}\to\infty$ (specifically, this is convergence in probability\cite{williams1991probability}).
We also expect a CLT for this quantity: as in Eq.~\ref{equ:clt-gen}, the distribution of the error $(\hat{A}_{\rm F}- \langle A\rangle_{\rm F})$ converges to a Gaussian with mean zero, and variance $\Sigma_{\TF}/M_{\rm F}$.
We will derive a formula for this variance, which will be compared later with the corresponding quantity for the three-level model.

For compact notation, it is convenient to define the recentred quantity of interest
\beq
    \dA(\CC) = A(\CC) - \langle A \rangle_{\rm F}.
\eeq
A significant contribution to $\Sigma_{\TF}$ comes from the randomness of the annealing procedure, this can be quantified as
\beq
    v(\CC) = \Var_{\rm J} \big[\Wn(\CC)\big]  
\label{equ:vC2}
\eeq
where the variance is again with respect to the annealing procedure (from coarse to fine).
Then, following Ref.~\onlinecite{kobayashi2019correction}, it can be shown that
\begin{equation}
    \Sigma_{\TF}
        = \langle \dA(\CC)^2 [w(\CC)^2+v(\CC)] \rangle_{\rm C}
    \label{equ:clt-Af}
\end{equation}
where $w(\CC)=\langle \Wn(\CC)\rangle_{\rm J} = p_{\rm F}(\CC)/p_{\rm C}(\CC)$, 
so one identifies $w(\CC)^2+v(\CC)$ as the mean square weight obtained from the annealing procedure.
Similarly, the estimator $\hat\Delta$ that appears in the difference estimate $\hat{A}_{\TF,\Delta}$ also obeys a CLT, with variance $\Var(\hat{\Delta}) \approx \Sigma_{\TF,\Delta}/M_{\TF}$, where
\begin{multline}
  \Sigma_{\TF,\Delta} = 
            \Big\langle \dA(\CC)^2 \big[w(\CC)^2+v(\CC)-1\big] \Big\rangle_{\rm C} 
            \\
            + 
            \Var_{\rm C}(A) -  \Var_{\rm F}(A)  \, .
               \label{equ:clt-Delta}
\end{multline}
As discussed in Ref.~\onlinecite{kobayashi2019correction}, if the computational cost of the coarse model is low then $M_{\TC}$ can be taken large enough that the variance of the coarse estimator $\hat{A}_{\TC}$ is negligible, in which case (\ref{equ:A-TML}) implies $\Var(\hat{A}_{\TF,\Delta}) \approx \Var(\hat{\Delta})$, and hence
\beq
    \Var(\hat{A}_{\TF,\Delta}) \approx
            \frac{1}{M_{\TF}} \Sigma_{\TF,\Delta} \; .
                        \label{equ:clt-Afml}
\eeq

Comparing (\ref{equ:clt-Af}) and (\ref{equ:clt-Delta}) -- which give the variances of $\hat{A}_{\TF}$ and $\hat{A}_{\TF,\Delta}$ respectively -- the term $v(\CC)$ in (\ref{equ:clt-Af}) is replaced by by $v(\CC)-1$ in \eqref{equ:clt-Delta}, which reduces the variance of the estimator.
We expect in general that $ \operatorname{Var}_{\rm C}(A)$ and $\operatorname{Var}_{\rm F}(A)$ should be similar in magnitude, in which case these terms in (\ref{equ:clt-Delta}) should have little effect.
Hence one expects that the estimator $\hat{A}_{\TF,\Delta}$ has lower variance than $\hat{A}_{\TF}$.
This is consistent with the results of Fig.~\ref{fig:ResultsStdDev}.

\subsection{Three-level method}

The results (\ref{equ:clt-Af},\ref{equ:clt-Afml}) are based on the property that each estimator is a sum of (nearly) independent random variables, which means that we can immediately apply standard Monte Carlo convergence results \cite{robert2004monte}.
This is not possible for the three-level method, since the resampling step correlates the configurations.
This makes the analysis of SMC-type algorithms challenging, but widely applicable results are available \cite{cappe2005inference,douc2008limit,chan2013general}.
The three-level method in Section \ref{sec:threelevel} is an implementation of a random-weight SMC method which has been analysed in Ref.~\onlinecite{rohrbach2022convergence}.

To analyse the variance of the three-level method, we require results analogous to (\ref{equ:clt-Af}), which depend on the mean square weights associated with the annealing procedure.  To this end, 
define the average of the final level weight
\begin{equation}
    w_2(\XX_2) = \langle \Wn_2(\XX_2) \rangle_{\rm J}
    \label{eqn:w2X}
\end{equation}
which 
fulfils \eqref{eqn:MarginalIntegralW2}. Similar to \eqref{equ:vC2}, the variance of this weight is
\beq
    v_2(\XX_2) = \Var_{\rm J} \big[ \Wn_{2}(\XX_2)\big].
    \label{eqn:v2X}
\eeq 
The averages in these equations are with respect to the second annealing step (from intermediate to fine level), starting at configuration $\XX_2$, see Sec.~\ref{sec:3-fine}.

For the contribution to the asymptotic variance of the first annealing step, it is important to consider a product of weight factors: $\Wn_1(\CC_1) w_2(\hat \XX_1)$.
The first factor in this product is the random weight $\Wn_1$ that is obtained by annealing from the coarse to the intermediate level, leading to the intermediate configuration is $\hat \XX_1 = (\CC_1, \hat \CF_1)$.
The second factor is the averaged weight $w_2(\hat \XX_1)$ from \eqref{eqn:w2X} associated with the second (subsequent) annealing step.
Combining \eqref{equ:jarz-fine-property-int} and \eqref{eqn:MarginalIntegralW2}, the average of the product is
\begin{equation}
    \langle \Wn_1(\CC_1) w_2(\hat \XX_1) \rangle_{\rm J} 
        = p_{\TF}(\CC_1) / p_{\TC}(\CC_1)
        = w(\CC_1)
\end{equation}
and the corresponding variance is
\beq
    v_1(\CC_1) = \Var_{\rm J} \big[\Wn_1(\CC_1) w_2(\hat \XX_1) \big]  \; .
    \label{equ:vi}
\eeq
Hence $w(\CC_1)^2+v_1(\CC_1)$ is the mean square value of $\Wn_1(\CC_1) w_2(\hat \XX_1)$ with respect to the the annealing process: this turns out to be a relevant quantity for the asymptotic variance.

The number of configurations $M_1, M_2$ can be varied between steps of the three-level method.
We formulate the asymptotic variance in the average number of configurations
\begin{equation}
    M_{\rm T} =\frac12(M_1 + M_2).
\end{equation}
If the two annealing steps have comparable cost, we can then directly compare the variances for different tapering rates at fixed $M_{\rm T}$.
Define also 
\begin{equation}
    c=\frac{M_{1}}{2M_{\rm T}}, \qquad \bar c = 1-c = \frac{M_{2}}{2 M_{\rm T}}.
\end{equation}
Then, a direct application of Theorem 2.1 of Ref.~\onlinecite{rohrbach2022convergence} gives a CLT for $\hat A_{\TF}^{\text{3L}}$: for large $M_{\rm T}$ we have
\beq
    \Var(\hat{A}_{\TF}^{\text{3L}}) 
        \approx \frac{1}{M_{\rm T}} \Sigma_{\TF} ^{\text{3L}}
    \label{equ:clt-Afml3L}
\eeq
with asymptotic variance
\beq
    \Sigma_{\TF} ^{\text{3L}}  = \frac{1}{2 c} \Sigma_{\rm F,1}^{\rm{3L}} +\frac{1}{2 \bar c}  \Sigma_{\rm F,2}^{\rm{3L}}
    \label{equ:SigTF3}
 \eeq
 with
\begin{align}
    \Sigma_{\rm F,1}^{\rm{3L}}  & =  \Big\langle \dA(\CC)^2 \big[w(\CC)^2+v_{1}(\CC)\big] \Big\rangle_{\rm C} \;, \nonumber
    \\
 \Sigma_{\rm F,2}^{\rm{3L}}  & =  \Big\langle \dA(\CC)^2 \big[w_{2}(\XX)^2+v_{2}(\XX)\big] \Big\rangle_{\rm I} \; .
 \label{equ:SigF-SigF}
\end{align}
The physical interpretation of these formulae will be discussed in the next subsection.  

Computing the asymptotic variance of the three-level difference estimator $\hat A_{\TF, \Delta}^{\text{3L}}$ is more difficult, since it involves difference of non-trivially correlated samples.
For some examples of multilevel difference estimators, upper bounds on the asymptotic variance have been developed in Refs.~\onlinecite{beskos2017multilevel, del2017multilevel}.
A detailed analysis of these bounds in the context of our algorithm is beyond the scope of this paper.

\subsection{Discussion of CLTs}

To understand the differences between the two- and three-level method, we compare the asymptotic variances of their corresponding final level estimators $\Sigma_{\TF}$ in \eqref{equ:clt-Af} and $\Sigma_{\TF}^{\text{3L}}$ in \eqref{equ:SigTF3}.
The variance of the three-level method has two contributions $\Sigma_{\TF,1}^{\rm 3L}$ and $\Sigma_{\TF,2}^{\rm 3L}$; they are the variances of two-level methods from the coarse to the fine model and the intermediate to the fine model, respectively.
The first term $\Sigma_{\TF,1}^{\text{3L}}$ is therefore directly related to $\Sigma_{\TF}$, where the variance of the importance weight $v(\CC)$ has been replaced by $v_1(\CC)$.

In order to make quantitative comparisons, we again consider the three-level method without intermediate resampling.
As discussed in Sec.~\ref{sec:variance}, this is a two-level method with a specific annealing process that consists of the concatenation of the two annealing processes of the three-level method.
For the concatenated annealing process, we have
\begin{equation}
    \Wn(\CC) =  \Wn_1(\CC) \Wn_2(\hat \XX),
\end{equation}
where $\hat \XX = (\CC, \hat \CF)$ is generated by the first annealing stage.
This means that
\begin{equation}
    v(\CC) = \Var_{\rm J} \Big[ \Wn_1(\CC) \Wn_2(\hat \XX) \Big],
    \label{eqn:VarConcatAnnealing}
\end{equation}
where the variance is now over the randomness of both annealing processes.
Comparing \eqref{eqn:VarConcatAnnealing} to \eqref{equ:vi}, we see that $v_1(\CC)$ computes the variance of the same importance weight, but after averaging over the second annealing stage in \eqref{eqn:w2X}.
We can apply Jensen's inequality\cite{williams1991probability} to show that
\begin{equation}
    v(\CC) \geq v_1(\CC) .
\end{equation}
By definitions (\ref{equ:clt-Af}, \ref{equ:SigTF3}), this directly implies
\begin{equation}
    \Sigma_{\TF,1}^{\text{3L}} \leq \Sigma_{\TF}.
\end{equation}

For the case without tapering $c = 1/2$, the three-level method therefore trades a reduction in the variance of the importance weights from coarse to fine in $\Sigma_{\TF,1}^{\rm 3L}$ for the addition of a term $\Sigma_{\TF,2}^{\rm 3L}$ that corresponds to the variance of a two-level method going from the intermediate to the fine level.
The possibility of tapering, i.e.~$c \neq 1/2$, further allows us to optimise the distribution of computation effort between the two stages, which is particularly useful if $\Sigma_{\TF,2}^{\rm 3L} \ll \Sigma_{\TF, 1}^{\rm 3L}$.

For our application to hard sphere mixture example in Sec.~\ref{sec:example}, the annealing process is computationally expensive and the resulting weights are noisy.
We are therefore in the situation where the variance $v(\CC)$ contributes substantially to the overall variance, and where we have constructed an intermediate in Sec.~\ref{sec:IntermediateLevel} that improves on the CG model.
Following the discussion above, this is the setting where we expect the three-level method to improve upon a two-level method, which is confirmed by the numerical results in Sec.~\ref{sec:NumericalResults}.
Further discussion of the effect of resampling on random-weight SMC methods can be found in Ref.~\onlinecite{rohrbach2022convergence}.

\section{Conclusions} \label{sec:conclusions}

We have introduced a three- and multilevel extension of the two-level simulation method first discussed in Ref.~\onlinecite{kobayashi2019correction}.
We have applied this method to a highly size-asymmetric binary hard-sphere system.
As shown in the numerical test in Section \ref{sec:NumericalResults} and theoretical results in Section \ref{sec:ConvergenceResults}, the introduction of intermediate resampling that distinguishes the two- from the three-level method can lead to substantial improvements in performance by reducing the variance in importance weights and by allowing efficient allocation of resources between levels via tapering.

\subsection{Hard sphere model}

In the application to binary hard-sphere systems, the introduction of an intermediate level required us to construct a semi-analytic estimate of the free energy of a system with partially inserted small particles.
For this, we have combined a highly accurate square-gradient theory with pre-computed ad hoc corrections, yielding an intermediate level that substantially improves the accuracy of the investigated quantities of interest compared to the initial coarse level.
Furthermore as we show in Appendix \ref{app:AdHocCorrection}, the three-level method appears robust with respect to slight deviations of the intermediate level.

Compared to our numerical example, Ref.~\onlinecite{kobayashi2021critical} applied the two-level method to larger and more dense systems than considered here, to investigate the critical point of demixing.
This was achieved by replacing the two-body CG model with RED potential used in this publication by a highly accurate two- and three-body potential.
The computation of accurate effective potentials entails a substantial upfront computational cost (compared to our construction of the intermediate level), but for the hard sphere mixtures this results in a CG level that is more accurate than our intermediate level.
Despite the challenges of keeping the variance of the importance weights under control for large systems, this turned out to be more efficient overall.

\subsection{Design principles for other potential applications}

We have emphasised throughout that our three-level modelling approach is generally applicable, whenever a suitable intermediate level can be constructed.  We can identify two main scenarios where this might be attempted.  
The first scenario is illustrated by the binary hard sphere mixture, which  is a two-scale system by construction (there are two species).  In this situation, there is no obvious intermediate level, and a careful construction is required, to design one.  Our results show that this strategy is possible -- it is worthwhile in this example because the system is very challenging to characterise by other methods, so the effort of constructing the intermediate level is worthwhile.

The second scenario -- where we
may expect a multilevel method to be particularly useful -- is that a multi-scale system admits a true hierarchy of coarse-grainings, such as
a system of long-chain polymers.
We can coarse-grain a polymer chain by representing groups of monomers by their centre of masses, with suitable effective interactions\cite{Pierleoni2007,d2015coarse}.
By varying the number of monomers per group, we get a hierarchy of CG models that could be targeted by a multilevel method.
For such methods to be efficient in such a scenario, we require high accuracy of the CG models, and an efficient annealing process to introduce the finer degrees of freedom analogous to the introduction of the small spheres in the hard sphere mixture.
Fulfilling these requirements is still challenging, and requires considerable physical insight about the specific polymer system of interest, but the hierarchical structure of the system hints that a suitable method might be fruitfully extended to more than three levels, with commensurately increased performance gains.
 
In both scenarios, careful thought is required to apply the three-level (or multi-level) methods: our approach is far from being a black-box method.  Still, the results presented here show that it can be applied in a practical (challenging) computational problem.

A separate limitation of multilevel methods is that the population of unique coarse configurations is fixed from the start, and  reduces with each subsequent resampling step.
This is closely related to the sample depletion effect commonly observed effect in particle filtering, and SMC methods in general~\cite{crisan2000convergence,doucet2009tutorial}.
For the multilevel method, we can address this by following each resampling step with a number of MCMC steps, to decorrelate duplicated configurations and further explore the system at the current level of coarse-graining\cite{crisan2000convergence}.
While such an approach is not feasible for the hard-sphere system where intermediate MCMC is limited by the cost of computing the required approximations, we expect this to be beneficial for example whenever intermediate physical systems are described in terms of effective, few-body interactions.

We end with a comment on the implementation of these methods.  The introduction of intermediate levels increases the complexity of the code required to simulate the systems.
It requires adding an intermediate stage to the annealing process and computing the required integrals, see Sec.~\ref{sec:IntermediateLevel}.
Additionally, when implementing the algorithm for the use on compute clusters, the resampling step requires the communication between all nodes.
However, we emphasise that while these extra steps require some extra programming, 
none of the additional steps of the three-level method have added significant computational cost in our example.

To conclude, our results show that the multilevel method can effectively make use of intermediate levels when available, leading to improvements in performance at fixed computational cost.
We look forward to further applications of multilevel methods in physical simulations.

\section*{Acknowledgements}
This work was supported by the Leverhulme Trust through research project Grant No. RPG–2017–203.

\section*{Author declarations}
\subsection*{Conflict of interest}
The authors have no conflicts to disclose.

\appendix

\section{Ensemble definitions, and estimation of partition functions} \label{app:FreeEnergy}

\subsection{Grand canonical ensemble} \label{app:DetailsModel}
We define the grand canonical ensemble of the hard sphere mixture discussed in Section \ref{sec:HSMixture}.
Recall that $k_{\rm B} T = 1$.
For the system of interest, the equilibrium average of a quantity of interest $A(\CC)$ in \eqref{eqn:FineAverage} is defined as
\begin{multline}
    \langle A \rangle_{\TF} 
        = \frac{1}{\Xi_{\TF}} \sum_{N=0}^{\infty} \sum_{n=0}^{\infty}
            \frac{e^{\muB N + \muS n}}{n! N! \sigB^{3N} \sigS^{3n}} \\
         \qquad \times \int A(\CC) e^{-U_{\TF}(\CC, \CF)} 
            \id \bfR_1 \cdots \id \bfR_{N}
            \id \bfr_1 \cdots \bfr_n,
        \label{eqn:defGCAverage}
\end{multline}
where each particle position is integrated over the periodic domain $[0, L]^3$.
For ease of notation, we introduce the integration measures $\id \CC, \id \CF$ that include the prefactors accounting for the indistinguishability of particles that appear in \eqref{eqn:defGCAverage}, which then becomes
\begin{equation}
    \langle A \rangle_{\TF} 
        = \frac{1}{\Xi_{\TF}} \int 
           A(\CC)  e^{\muB N + \muS n - U_{\TF}(\CC, \CF) }
          \id \CC \id \CF ,
\end{equation}
consistent with (\ref{equ:pf},\ref{eqn:FineAverage}).
By definition, we require that $p_{\TF}$ is normalised as $\int p_{\TF}(\CC, \CF) \id \CC \id \CF = 1$, so we have
\begin{multline}
    \Xi_{\TF} 
        = \sum_{N=0}^{\infty} \sum_{n=0}^{\infty}
            \frac{e^{\muB N + \muS n}}{n! N! \sigB^{3N} \sigS^{3n}} \\
            \times \int e^{-U_{\TF}(\CC, \CF)} 
                \id \bfR_1 \cdots \id \bfR_{N}
                \id \bfr_1 \cdots \bfr_n.
\end{multline}
The relevant quantities of the CG model $\langle A \rangle_{\TC}$ are defined analogously.

\subsection{Estimation of the partition function} \label{app:AIS}
The implementation of the two- and three-level methods requires the computation of the small-particle partition function that appear in the importance weights.
We use a method based on Jarzynski's equality \cite{jarzynski1997nonequilibrium,crooks2000path} that yields an unbiased estimator, see also Ref.~\onlinecite{kobayashi2019correction}.
In the statistics literature, this is also known as Annealed Importance Sampling \cite{neal2001annealed}.
We first give a short summary of the method in App.~\ref{app:AISTheory} and then discuss how the annealing processes are implemented for the two- and three-level method in App.~\ref{app:AISApplication}.
The parameters used for the numerical tests are given in App.~\ref{app:AISParameters}.

\subsubsection{Theoretical details}\label{app:AISTheory}
We derive an annealing process that inserts the small particles $\CF$ for a fixed big particle configuration $\CC$.  This process produces weighted configurations that correctly characterise the FG distribution.
We closely follow the results from Appendix A of Ref.~\onlinecite{kobayashi2019correction}, see also Refs.~\onlinecite{neal2001annealed,jarzynski1997nonequilibrium,crooks2000path,oberhofer2009efficient}.

Let $p_{\text{s}}(\CC, \CF)$ and $p_{\text{e}}(\CC, \CF)$ be two probability distributions for the FG model of the form
\begin{equation}
    p_{\alpha}(\CC, \CF) = \frac{1}{\Xi_\alpha} e^{\Phi_\alpha(\CC, \CF)}, \;\; \alpha \in \{\text{s}, \text{e}\}.
\end{equation}
The corresponding marginal distributions are $p_{\alpha}(\CC) = \int p_{\alpha}(\CC, \CF) \id \CF$.
The distributions $p_{\text{s}}, p_{\text{e}}$ are the start and end point of an annealing process, with a sequence of intermediate distributions
\begin{equation}
    p_{k}(\CC, \CF) =  \frac{1}{\Xi_k} e^{\Phi_k(\CC, \CF)}, \;\; k=0, \dots, K,
\end{equation}
where $p_0 = p_{\text{s}}$ and $p_{K} = p_{\text{e}}$.

Let $\CC$ be a sample from $p_{\text{s}}(\CC)$: this configuration remains fixed during the annealing process.
We anneal the small particles, as follows:
first sample an initial small particle configuration $\CF$ from $p_{\text{s}}(\CF \mid \CC)$, the conditional distribution of $p_{\text{s}}$.
This distribution is $p_0$ so write  $\CF_0 = \CF$ and set $k=1$: then apply a sequence of MC steps with transition kernel $q_{\CC, k}(\CF_{k-1} \to \CF_{k})$ that is in detailed balance with the small particle distribution $p_{k}(\CF \mid \CC)$.  Iterate this process for  $k=1, \dots, K-1$: this yields a sequence of small-particle configurations $(\CF_0, \CF_1, \dots, \CF_{K-1})$.
The big-particle configuration $\CC$ stays fixed throughout this process.

The relevant results of this procedure are the final small-particle configuration $\hat \CF = \CF_{K-1}$ and an annealing weight 
\begin{equation}
    \hat W_{\TA} = e^{\sum_{k=1}^{K} \left[\Phi_k(\CC, \CF_{k-1}) - \Phi_{k-1}(\CC, \CF_{k-1}) \right]}.
\end{equation}
Given the initial coarse configuration $(\CC, \CF)$, the MC steps define a probability distribution over the weight $\hat W_{\TA}$ and the final small particle configuration $\hat \CF$, which we denote by
\begin{equation}
    \kk(\hat W_{\TA}, \hat \CF \mid \CC, \CF).
\end{equation}
Given the initial configuration $(\CC, \CF)$, averages with respect to the annealing process are denoted by $\langle \, \cdot \, \rangle_{\rm J}$.

We now show that this annealing process produces weighted samples of $p_{\text{e}}$, up to a constant.
More specifically:
\begin{multline}
    \int \hat W_{\TA} \kk(\hat W_{\TA}, \hat \CF \mid \CC, \CF) p_{\text{s}}(\CF \mid \CC) \id \hat W_{\TA} \id \CF \\
        = \frac{\Xi_{\text{e}}}{\Xi_{\text{s}}} \frac{p_{\text{e}}(\CC, \hat \CF)}{p_{\text{s}}(\CC)} . \label{eqn:AISAverage}
\end{multline}
This implies that averaging over the start distribution $p_{\text{s}}$ and the annealing process yields
\begin{equation}
   \langle \hat W_{\TA} B(\CC, \hat \CF) \rangle_{\text{J}, p_{\text{s}}}
        = \frac{\Xi_{\text{e}}}{\Xi_{\text{s}}} \langle B(\CC, \CF) \rangle_{p_{\text{e}}}
        \label{eqn:AISAverageB}
\end{equation}
for any function $B = B(\CC, \CF)$, which may depend on both big and small particles.

To show \eqref{eqn:AISAverage}, we compute the average over the annealing process explicitly
\begin{multline}
    \int \hat W_{\TA} \kk(\hat W_{\TA}, \hat \CF \mid \CC, \CF) p_{\text{s}}(\CF \mid \CC) \id \hat W_{\TA} \id \CF \\
    = \int e^{\sum_{k=1}^{K} \left( \Phi_k(\CC, \CF_{k-1}) - \Phi_{k-1}(\CC, \CF_{k-1})\right)} \prod_{k=1}^{K-1} q_{\CC, k}(\CF_{k-1} \to \CF_{k}) \\
        \times p_{\text{s}}(\CF_{0} \mid \CC) \id \CF_0 \cdots \id \CF_{K-2}. 
    \label{eqn:IntAIS1}
\end{multline}
By rearranging the factors in the exponential, the right-hand side of \eqref{eqn:IntAIS1} becomes
\begin{multline}
    \int \prod_{k=1}^{K-1} q_{\CC, k}(\CF_{k-1} \to \CF_{k}) 
        e^{\Phi_k(\CC, \CF_{k-1}) - \Phi_{k}(\CC, \CF_{k})}  
        \\
        \times e^{\Phi_{K}(\CC, \CF_{K-1})} \frac{p_{\text{s}}(\CF_0 \mid \CC)}{e^{\Phi_{0}(\CC, \CF_{0})}}
            \id \CF_{0} \cdots \id \CF_{K-2} .
        \label{eqn:IntAIS2}
\end{multline}
Detailed balance of the Markov kernels $q_{\CC, k}$ implies
\begin{equation}
    q_{\CC, k}(\CF_{k-1} \to \CF_{k}) 
        e^{\Phi_k(\CC, \CF_{k-1}) - \Phi_{k}(\CC, \CF_{k})} 
        = q_{\CC, k}(\CF_{k} \to \CF_{k-1}), \label{eqn:DetailedBalance}
\end{equation}
and by definition
\begin{equation}
    e^{\Phi_{K}(\CC, \CF_{K-1})} \frac{p_{\text{s}}(\CF_0 \mid \CC)}{e^{\Phi_{0}(\CC, \CF_{0})}}
        = \frac{\Xi_{\text{e}}}{\Xi_{\text{s}}} \frac{p_{\text{e}}(\CC, \CF_{K-1})}{p_{\text{s}}(\CC)}. \label{eqn:IntAISCond}
\end{equation}
Using \eqref{eqn:DetailedBalance} and \eqref{eqn:IntAISCond}, Eq.~\ref{eqn:IntAIS2} simplifies to
\begin{multline}
    \int \prod_{k=1}^{K-1} q_{\CC, k}(\CF_{k} \to \CF_{k-1}) \id \CF_{0} \cdots \id \CF_{K-2}
        \\
         \times \frac{\Xi_{\text{e}}}{\Xi_{\text{s}}} \frac{p_{\text{e}}(\CC, \CF_{K-1})}{p_{\text{s}}(\CC)} .
        \label{eqn:IntAIS3}
\end{multline}
Since $q_{\CC, k}(\CF_{k} \to \CF_{k-1})$ is a normalised probability density for $\CF_{k-1}$, we can perform the integrals in \eqref{eqn:IntAIS3} one by one, yielding \eqref{eqn:AISAverage}.

\subsubsection{Application to the two- and three-level method}\label{app:AISApplication}
This section describes the details of the annealing processes used in the two- and three-level method.
We first discuss its implementation for the two-level method, before showing how to split this process into two stages for the three-level method.

The two-level method starts with samples $\CC$ of the CG model $p_{\TC}$.
We describe the annealing process~\cite{kobayashi2019correction} which produces a weight $\hat W(\CC)$ and small particle configuration $\hat \CF$ that fulfils \eqref{equ:jarz-fine-property}.
Since we have no initial small particle distribution, we cannot directly apply the results of App.~\ref{app:AISTheory} and need to proceed in two steps.
Let
\begin{equation}
    p_{\mu_S}(\CF \mid \CC) = p_{\TF}(\CF \mid \CC) =  \frac{1}{\Xi[\CC, \mu_S]} e^{\mu_S n - U_{\TF}(\CC, \CF)}
    \label{eqn:DefPCMuS}
\end{equation}
be the distribution of small particles around a fixed big-particle configuration $\CC$, where we now explicitly note the dependence on the small particle chemical potential $\mu_S$.
Computing the unnormalised importance weight $\hat W(\CC)$ from \eqref{eqn:defUnnormalisedWeight} requires an estimate of the partition function of the small particles $\Xi[\CC, \mu_S]$.
For a system with a small value of the chemical potential $\mu_0 \ll \mu_S$, we can directly estimate this quantity
\begin{equation} \label{eqn:mu0PartitionFunction}
    \Xi[\CC, \mu_0] = \frac{1}{\mathbb{P}_{p_{\mu_0}(\cdot \mid \CC)}\left(n=0\right)}
\end{equation}
as it is the reciprocal probability of having zero small particles in a system with fixed $\CC$.
For small enough $\mu_0$, this value is close to $1$ and can be estimated quickly by a GCMC simulation that decorrelates quickly due to the low density of small particles.
Since we can compute this value to a very low variance at negligible cost, we consider our estimate of $\Xi[\CC, \mu_0]$ it to be exact and we neglect the influence of its fluctuations on the overall variance of the method.
Furthermore, we assume that we can generate samples from the low chemical potential distribution of small particles $p_{\mu_0}(\CF \mid \CC)$.

Starting with a sample $\CF$  of the initial small particle distribution $p_{\mu_0}(\CF \mid \CC)$, we can now apply steps of the annealing process defined in the previous section.
We define the steps of the annealing process by slowly increasing the chemical potential of the small particles $\mu_k$ in $K+1$ steps, from $\mu_0$ to $\mu_K = \muS$ while keeping the CG distribution fixed.
More specifically, we simulate an annealing process for the sequence of probability distributions
\begin{equation}
    p_{\TC, \mu_k}(\CC, \CF) = p_{\TC}(\CC) p_{\mu_k}(\CF \mid \CC), \;\; k=0, \dots, K, 
    \label{eqn:DefPMuL}
\end{equation}
yielding an annealing weight $\hat W_{\text{A}}$ and a fine-particle configuration $\hat \CF$ as described previously.

Averaging over the initial distribution of small particles and the annealing process and using (\ref{eqn:AISAverage},\ref{eqn:DefPCMuS},\ref{eqn:DefPMuL}) yields
\begin{multline}
    \int \hat W_{\TA} \kk(\hat W_{\TA}, \hat \CF \mid \CC, \CF) p_{\mu_0}(\CF \mid \CC) \id \hat W_{\TA} \id \CF \\
        = \frac{\Xi[\muS, \CC]}{\Xi[\mu_0, \CC]} p_{\TF}(\hat \CF \mid \CC) . 
    \label{eqn:AISAverage2L}
\end{multline}
Combining this with \eqref{eqn:GCPotential}, we have
\begin{equation}
    \frac{\Xi[\muS, \CC]}{\Xi[\mu_0, \CC]} p_{\TF}(\hat \CF \mid \CC)
         = \frac{\Xi_{\TF}}{\Xi_{\TC}} \frac{p_{\TF}(\CC, \hat \CF)}{p_{\TC}(\CC)} \frac{e^{-U_{\TC}(\CC)}}{\Xi[\mu_0, \CC]}. 
\end{equation}
Thus, we scale the weight that is produced by the annealing process
\begin{equation}
    \hat W(\CC) = \Xi[\mu_0, \CC] e^{U_{\TC}(\CC)}  \hat W_{\TA}.
\end{equation}
For this weight, the annealing process fulfils \eqref{equ:jarz-fine-property} when we include the sampling from the distribution of the initial small particles $\CF \sim p_{\mu_0}(\CF \mid \CC)$ as part of the annealing process.

For the three-level method, we split the annealing process discussed above into two consecutive steps.
The first part follows exactly the same steps as above, where the annealing process increases the chemical potential $\mu_k$ from a small value $\mu_0$ to $\mu_S$.
The only difference is that we include the potential $\tilde{U}$ of the intermediate distribution: in place of \eqref{eqn:DefPCMuS} we have
\begin{equation}
    p_{\mu_S}(\CF \mid \CC)  = p_{\TI}(\CF \mid \CC)
\end{equation}
so that small particle insertion is suppressed in regions far from large particles.
As before, the annealing process results in a (scaled) weight $\hat W_1(\CC)$ and small particle configuration $\hat \CF_1$ that now fulfils \eqref{equ:jarz-fine-property-int}.

For the second step of the three-level method, we need to define an annealing process that fulfils \eqref{equ:WF-ave}.
We start with a sample $\XX_2 = (\CC_2, \CF_2)$ from the intermediate level $p_{\TI}$.
Since this configuration already contains small particles, we can directly apply the results of App.~\ref{app:AISTheory} to anneal from $p_{\TI}(\CC, \CF)$ to $p_{\TF}(\CC, \CF)$.
This is achieved by a sequence of intermediate annealing distributions that increase the parameter $\delta$ of the potential \eqref{eqn:CosinePotential}, so that the volume available to the small particles is slowly increased.
This is done in $K+1$ steps from the parameter $\delta_0 = \delta_{\rm free}$ (the intermediate level) to a final value
\begin{equation}
    \delta_{K} = \max_{\bfr} \dist(\bfr, \CC_2),
\end{equation}
at which point the suppression potential does not affect any point in the domain.
Then, the intermediate level with $\delta_{K}$ corresponds to the fine-level distribution, up to the correction factor $e^{\Phi_{\text{corr}}(\CC)}$ in \eqref{equ:int-cor} that only depends on the big particles.
Following from \eqref{eqn:AISAverage}, this annealing process with scaled weight $\hat W_2(\CC) = \hat W_{\rm A} e^{-\Phi_{\rm corr}(\CC)}$ and final small particle configuration $\hat \CF_2$ fulfils the property \eqref{equ:WF-ave}.

\subsubsection{Annealing schedules for simulations}\label{app:AISParameters}

The importance weights produced by the annealing process are unbiased, and this feature is independent of the details of the annealing schedules.  In this sense, the algorithm is valid for any schedule.  
However, the variance of the computed importance weights depends strongly  on the choice of schedule.

The initial chemical potential $\mu_0$ is chosen such in a system with no big particles, there would be an average of $\bar n_0 = 0.01$ small particles present, that is the initial reservoir volume fraction is $\eta_{S,0}^{r} = 0.01/L^3$.

For the first stage annealing process, we increase the chemical potential in steps $\Delta \mu_k$ such that the average change in the number of small particles would be $\delta \bar n_k = 0.2$, in a system where no big particles were present.
For the second stage, we increase $\delta_k$ in fixed steps $\Delta\delta_k = \sigS/20000$.
In both cases, we run one GCMC sweep between each step.

To compute accurate FG reference results, used for example in Figs.~\ref{fig:Weights} and \ref{fig:Convergence}, we apply the two-level method using the same annealing strategy as for the first stage of the three-level method but with $\delta \bar n_k = 0.05$, for increased accuracy.
Note that for the numerical tests in Figs.~\ref{fig:ResultsStdDev} and \ref{fig:ResultsStdDevAlpha} that directly compare the performance of the two- and three-level method, the two-level method uses the same annealing schedule as the three-level method outlined above.
The only difference is the lack of resampling.

\section{Details of the intermediate level} \label{app:DetailsIntermediateLevel}
\subsection{Perturbative approximation of non-homogeneous hard-sphere fluid}\label{app:DetailsApprox}

This section derives (\ref{equ:gam}) of the main text.
To this end, consider a homogeneous hard sphere fluid at chemical potential $\mu$ and add a perturbing potential 
\newcommand{\bfq}{\bm{q}}
\begin{equation}
    \Egen(\bfr) = a \sin(\bfq \cdot \bfr)
\end{equation}
in \eqref{eqn:PerturbedSmallSphereFluid}.
In a finite periodic system then $\bfq$ should be a reciprocal lattice vector.
We aim to estimate the free energy difference between the perturbed and homogeneous system.
For this, we follow the steps of the local density approximation discussed in Ref.~\onlinecite{evans1979nature}, see also Chapter 6 of \onlinecite{hansen2013theory}.
We approximate this difference as
\begin{align}\label{eqn:sgdiff}
    \delta\Phi[\Egen] &= \Phi - \Phi[\Egen] \nonumber \\
        & \approx \int \press(\mu - \Egen) - \press(\mu) + \gam(\mu -  \Egen) |\nabla \Egen|^2 \id \bfr.
\end{align}
To compute $\gam$, we need to investigate both sides of equation \eqref{eqn:sgdiff}.
Starting with the right-hand side,
we assume that $\gam$ is smooth, therefore we can approximate it for small $a$ by a constant $\gam(\mu - \Egen) \equiv \gam(\mu)$.
To compute the integral of the pressure difference in \eqref{eqn:sgdiff}, we expand around $\mu$
\begin{align}
    \int  \press(\mu - &\Egen) - \press(\mu) \id \bfr \nonumber \\
         &= \int -\press'(\mu) \Egen(\bfr) + \frac{\press''(\mu)}{2} \Egen(\bfr)^2 \id \bfr + \mathcal{O}(a^3) \nonumber \\
         &= \frac{V}{4} a^2 \press''(\mu) + \mathcal{O}(a^3)
\end{align}
and integrate the gradient correction term
\begin{align}
    \int \gam(\mu) |\nabla \Egen|^2 \id \bfr
        &= q^2 a^2 \gam(\mu) \int \cos^2(\bfq \cdot \bfr) \id \bfr \nonumber \\
        &= \frac{V}{2} q^2 a^2 \gam(\mu).
\end{align}
Overall, we obtain
\begin{multline} \label{eqn:IntegralApproxDeltaPhi}
    \int \press(\mu - \Egen) - \press(\mu) + \gam(\mu -  \Egen) |\nabla  \Egen|^2 \id \bfr \\
         \approx \frac{V}{2} a^2 \left( q^2 \gam(\mu) + \frac{1}{2} \press''(\mu)\right).
\end{multline}

\begin{figure*}
    \includegraphics[width=\textwidth]{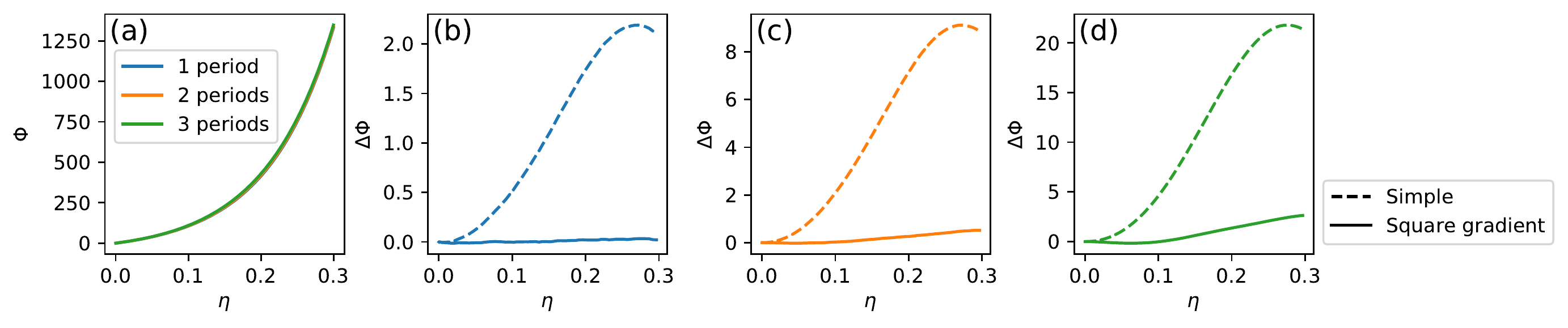}
    \caption{
        A numerical test of the approximation accuracy of the square-gradient method for computing free energies of non-homogeneous hard-sphere fluids.
        (a) The free energy $\Phi$ of the hard-sphere fluid described in Appendix \ref{app:exampleSG} with a cosine potential with $1, 2$, and $3$ periods within the simulation box as a function of the volume fraction $\eta$.
        (b,c,d) The difference between the free energy and its pressure integral approximation in \eqref{eqn:sgapprox} for a cosine potential with one (b), two (c) or three (d) periods, with (solid line) and without (dashed lines) the square-gradient term.
    }
    \label{fig:SGApproxAccuracy}
\end{figure*}

To investigate the left-hand side of \eqref{eqn:sgdiff}, we look at perturbations of the free energy in the limit of small $a$.
We can express the derivatives of the free energy in $a$ as equilibrium averages of the perturbed system.  For $a=0$, 
\begin{align}
    \frac{\partial \Phi[\Egen]}{\partial a} 
        &= - \left\langle \rho_{\bfq} 
        \right\rangle_{\mu} = 0 , \\
    \frac{\partial^2 \Phi[\Egen]}{\partial a^2} & = - \left\langle |\rho_{\bfq}|^2 \right\rangle_{\mu} ,
\end{align}
with $\rho_{\bfq} = \sum_{j=1}^n \sin(\bfq\cdot\bfr_j)$.
The first average is zero by translational invariance.
For the second average, use that $\langle |\rho_{\bfq}|^2 \rangle_\mu = \langle n \rangle_\mu S(\mu;q)/2$ where $S(\mu;q)$ is the structure factor of the fluid~\cite{hansen2013theory} at chemical potential $\mu$.  So
\begin{equation}
    \frac{\partial^2 \Phi[\Egen]}{\partial a^2} 
        = - \frac{\langle n \rangle_{\mu}}{2} S(\mu;q)
        = - \frac{3V}{\pi \sigma_S^3} \eta S(\mu;q). \numberthis \label{eqn:ddaPerturbation}
\end{equation}
%
Now differentiate \eqref{eqn:sgdiff} twice with respect to $a$, yielding
\begin{equation}
        \frac{3V}{\pi \sigma_S^3} \eta S(\mu;q) \approx V \left[ q^2 \gam(\mu) + \frac{1}{2} \press''(\mu) \right]
\end{equation}
where the left hand side used (\ref{eqn:ddaPerturbation}) and the right hand side was approximated with (\ref{eqn:IntegralApproxDeltaPhi}) before differentiating.

Finally, differentiate with respect to $q$ and send $|q| \to 0$: we can identify the second order term of the square-gradient approximation as
\begin{equation}
    \gam(\mu) = \frac{3\eta}{2\pi \sigma_S^3}  \frac{\partial^2}{\partial q^2} S(\mu; q) \Big\rvert_{q=0}, 
\end{equation}
which is (\ref{equ:gam}).

\subsection{Accuracy of the square-gradient approximation} \label{app:exampleSG}

The intermediate level that is constructed in Section \ref{sec:IntermediateLevel} relies on 
the approximation (\ref{eqn:sgapprox}) for
 the free energy of a non-homogeneous hard-sphere fluid.
This section discusses the accuracy of this approximation for a system that only contains small particles, in an external potential.

We consider a grand-canonical ensemble of small hard-spheres ($\sigma_{S} = 1$) in a periodic box $V = [0, L]^3$ of length $L=10\sigma_{S}$ without any big particles.
We perturb this system by a one-dimensional cosine potential with $m$ periods
\begin{equation}
    \Egen_{m, \cos}(r) = 2 \left[ \cos\left(\frac{2 m \pi}{10} r\right) + 1 \right].
\end{equation}
We apply this potential to the first component $r_1$ of the small particle positions $\bfr = (r_1, r_2, r_3)$.
The potential $\Egen_{m, \cos}$ has a fixed maximal strength of $4$.
We vary the steepness of the potential by varying the number of periods of the cosine $m=1, 2, 3$; as $m$ increases, the potential changes more rapidly.
We expect this to make our approximations increasingly inaccurate, as it is constructed under the assumption that the derivatives of the external potential are small.

We have computed the predicted free energy \eqref{eqn:sgapprox} in this system, as well as the cruder approximation \eqref{eqn:eqintegral}, which lacks the square gradient approximation.
We compare these values with the true free energy, computed via thermodynamic integration, and investigate the dependence of the accuracy of the approximation on the small particle volume fraction $\etaS$ and the steepness of the external potential.
The result of this computation are shown in Figure \ref{fig:SGApproxAccuracy}.

Figure \ref{fig:SGApproxAccuracy}(a) shows that the absolute value of the free energy depends weakly on the number of periods $m$ in the external potential.
As expected, the error of the approximation methods increases substantially in Figures \ref{fig:SGApproxAccuracy}(b)--(d), as the number of periods $m$ increases, as does the steepness of the cosine potential $\Egen_{m, \cos}$ . 
The absolute prediction error also increases in the volume fraction $\eta_{S}^r$.
For all $m$ considered here, the square-gradient method \eqref{eqn:sgapprox} substantially outperforms the prediction \eqref{eqn:eqintegral}.
For $m=1$, where the external potential varies the slowest, the square-gradient approximation is nearly exact, which confirms the accuracy of the square-gradient factor $\gam = \gam(\mu)$ in Appendix \ref{app:DetailsApprox}.
In addition to the increasing error on increasing $m$, the relative improvement of the square-gradient method over the simple approximation decreases.  This indicates that higher order terms in the derivative of the potential start to become more important.

The choice of the $1d$ cosine potential $\Egen_{m, \text{cos}}$ here is motivated by the use of a half-period of the cosine to introduce the suppression potential in the construction of the intermediate level in Section \ref{sec:IntermediateLevel}.
For the numerical examples in Section \ref{sec:NumericalResults}, the suppression potential has a (half-period) length of $l=3.5$ and a maximal strength of $s = 4.4$. 
In terms of the maximal squared gradient that appears in this potential, it lies between the cases $m=1$ and $m=2$.
This indicates that our square-gradient approximation is appropriate for the use of predicting the free energy of the partially inserted system in the binary hard-sphere example.

\section{Influence of the ad hoc correction factor} \label{app:AdHocCorrection}

\begin{figure}
    \centering
    \includegraphics[width=\columnwidth]{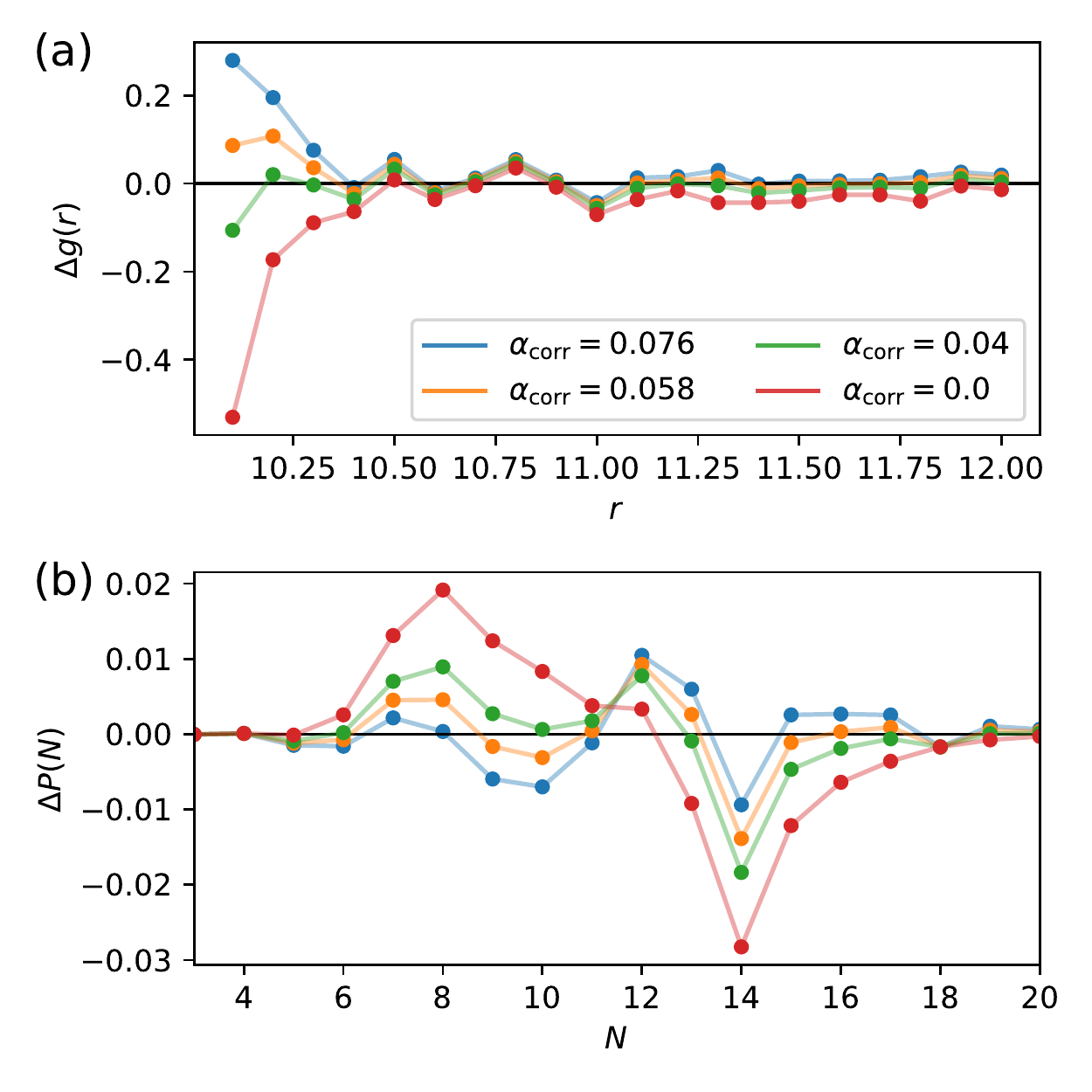}
    \caption{
        The difference between the estimates at the intermediate and final level for (a) the pair correlation function $g(r)$ and (b) the distribution of the number of big particles for different values of the ad-hoc correction factor $\alpha_{\text{corr}}$.
    }
    \label{fig:ExamplePropertiesAlpha}
\end{figure}

The construction of the intermediate level in Section \ref{sec:IntermediateLevel} included an ad hoc correction term that was identified using preliminary computations.
In this section, we take another look at this parameter and discuss its importance for the performance of the three-level method.
For this, we consider four values for the correction factor
\begin{equation} \label{eqn:ValuesAlpha}
    \alpha_{\text{corr}} \in \left\{ 0, 0.04, 0.058, 0.076 \right\}    
\end{equation}
between $\alpha_{\text{corr}} = 0$ (no ad hoc correction, which means $p_{\TI} = p_{\TI}^{(2)}$), and $\alpha_{\text{corr}} = 0.076$ (corresponding to a linear least-square fit to the weights in Figure \ref{fig:AdHocCorrection}).
The effect of the choice of correction factor is displayed in Figure \ref{fig:ExamplePropertiesAlpha} where we show the differences between the intermediate levels and the FG estimate for the two quantities of interest (see Figure \ref{fig:ExampleProperties} for their FG averages).
The ad-hoc correction has a noticeable influence on the quantities of interest, especially for the pair correlation function $g(r)$ when the two particles are almost touching ($r \approx 10$).  This is in the relevant region for the coordination number $N_c$, which was measured as part of the performance test in Section \ref{sec:variance}.

\begin{figure}
    \centering
    \includegraphics[width=\columnwidth]{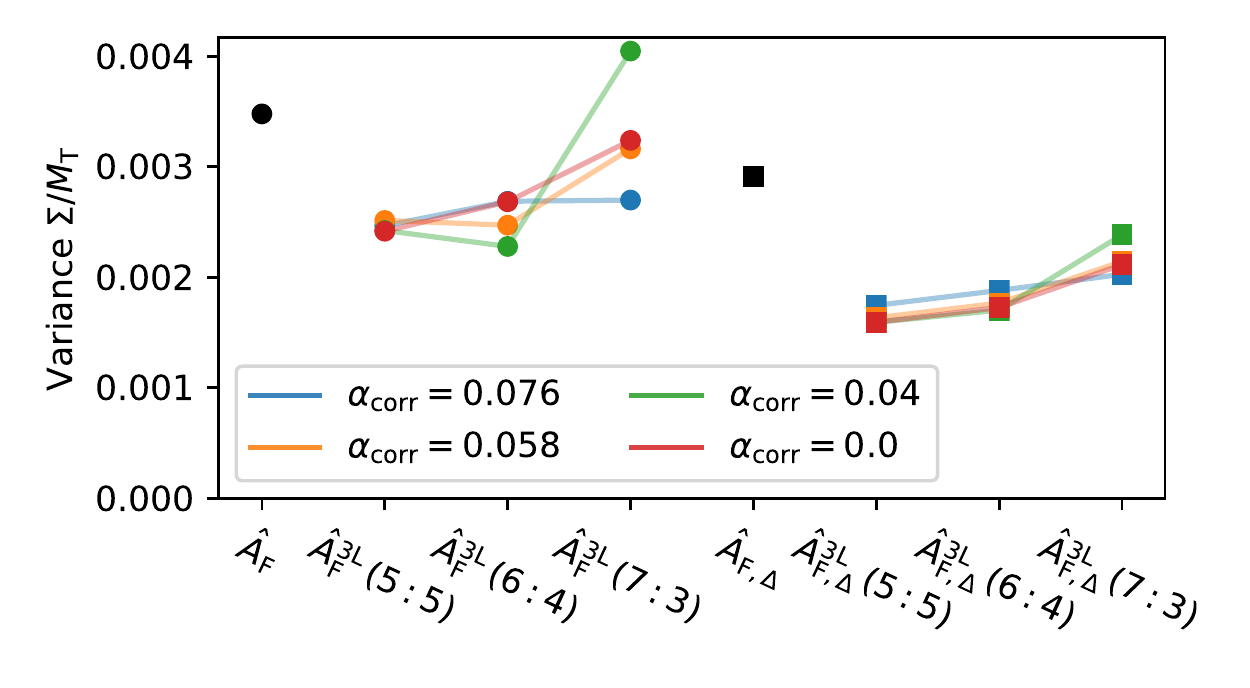}
    \caption{
        The sample variance of 60 independent estimates of the coordination number $N_c$, for different values of the ad hoc correction factor $\alpha_{\text{corr}}$ (see also Figure \ref{fig:ResultsStdDev}.)
    }
    \label{fig:ResultsStdDevAlpha}
\end{figure}

To determine how much the three-level method depends on $\alpha_{\text{corr}}$, we repeated this performance test for the different values in \eqref{eqn:ValuesAlpha}.
The measured sample variances for $N_{\rm runs}=60$ independent realisations of the estimators are shown in Figure \ref{fig:ResultsStdDevAlpha}.
As before, the results for different $\alpha_{\rm corr}$ are highly correlated, because they share the same configurations.
The (bootstrap) standard errors of the values in Figure \ref{fig:ResultsStdDevAlpha} are comparable to the ones shown in Figure \ref{fig:ResultsStdDev}; the same caveats apply and we have omitted them here for clarity of presentation.

The main takeaway from Figure \ref{fig:ResultsStdDevAlpha} is that for our example, the three-level estimators without tapering outperform the corresponding two-level estimators, independent of the choice of $\alpha_{\text{corr}}$.  That is,
the method appears to be robust with respect to modifications of the intermediate level, even if the mean quantity of interest differs significantly between intermediate distributions.

Given the low number of samples, the exact variance figures should not be over-interpreted.
Nevertheless, the trends in Figure \ref{fig:ResultsStdDevAlpha} illustrate two aspects of the SMC methodology that are relevant to applications.
First, the details of the intermediate level become more important when we increase the tapering rate, as one can for example see by comparing the four different values for the final-level estimator $\hat A_{\TF}^{\text{3L}}$ without tapering and with $7 : 3$ tapering.
Secondly, this effect is dampened for the difference estimators.
The general robustness of the difference estimator in the example considered here supports our assertion that for appropriately defined levels, it should be the preferred estimator when applying the three-level method in practice.

\bibliography{references,extra}

\end{document}